\documentclass[prb,twocolumn,amsmath,amssymb,floatfix]{revtex4}
\usepackage{graphicx}
\usepackage{bm}

\usepackage{color}

\newcommand{\be}{\begin{equation}}
\newcommand{\ee}{\end{equation}}
\newcommand{\beq}{\begin{eqnarray}}
\newcommand{\eeq}{\end{eqnarray}}
\newcommand{\nn}{\nonumber}

\newcommand{\bra}{\langle}
\newcommand{\ket}{\rangle}
\newcommand{\sgn}{\mathrm{sgn}}

\newcommand{\one}{1}
\newcommand{\two}{2}
\newcommand{\three}{3}

\newcommand{\eref}[1]{Eq.~\ref{#1}}
\newcommand{\sref}[1]{Sec.~\ref{#1}}
\newcommand{\fref}[1]{Fig.~\ref{#1}}

\newcommand{\cref}[1]{Chp.~\ref{#1}}
\newcommand{\aref}[1]{Appendix~\ref{#1}}

\newcommand{\rcite}[1]{Ref.~\onlinecite{#1}}

\begin{document}

\title{Classifying and measuring the geometry of the quantum ground state manifold}
\author{Michael Kolodrubetz$^\ast$, Vladimir Gritsev$^\dagger$, Anatoli Polkovnikov$^\ast$}
\affiliation{$^\ast$Physics Department, Boston University, 590 Commonwealth Ave., Boston, MA 02215, USA,\\
$^\dagger$ Department of Physics, University of Fribourg, Chemin du Musee 3, 1700 Fribourg, Switzerland}

\begin{abstract}

From the Aharonov-Bohm effect to general relativity, geometry plays a central role in modern physics.   
In quantum mechanics many physical processes depend on the Berry curvature. However, 
recent advances in quantum information theory have highlighted the role of its symmetric counterpart, the 
quantum metric tensor.  In this paper, we perform a detailed analysis of the ground state 
Riemannian geometry induced by the metric tensor, using the quantum XY chain in a transverse field as our
primary example.  We focus on a particular geometric invariant -- the Gaussian curvature -- 
and show how both integrals of the curvature within a given phase and singularities of the 
curvature near phase transitions are protected by critical scaling theory.  For cases where
the curvature is integrable, we show that the integrated curvature provides a new geometric
invariant, which like the Chern number characterizes individual phases of matter.  For
cases where the curvature is singular, we classify three types -- integrable,
conical, and curvature singularities -- and detail situations where each type of singularity
should arise.  Finally, to connect this abstract geometry to experiment, we discuss three different
methods for measuring the metric tensor, namely via integrating a properly weighted noise spectral function
and by using leading order responses of the work distribution to ramps and quenches in quantum many-body systems.

\end{abstract}

\maketitle


Understanding the geometry and topology of quantum ground state manifolds
is a key component of modern many-body physics.  For example,
it has become standard practice to characterize topological phases by their Chern number, 
defined as the integral of Berry curvature over a closed manifold in parameter space. 
Examples of this include the quantum Hall effect \cite{Thouless1982_1,Niu1985_1}, 
topological insulators \cite{Kane2005_1,Sheng2006_1}, 
integer and half integer spin chains \cite{Berry1984_1,Canali2003_1},  and many others. 
Non-zero Berry curvature is typically associated with broken time reversal symmetry, 
either explicitly by external coupling to a 
time reversal breaking field or implicitly by splitting the ground state manifold 
into different sectors, each of which breaks time reversal symmetry\cite{Sheng2006_1,Fu2006_1}.

However, in addition to the Berry curvature, which describes the flux of 
Berry phase within the ground state manifold, 
another important quantity is its symmetric
counterpart -- the quantum (Fubini-Study\footnote{Note that the metric tensor
we describe is really only locally equivalent to the Fubini-Study metric.  This 
comes from the fact that our metric operates on a relatively small manifold
of physical control parameters, whereas the Fubini-Study metric is formally
defined as a distance within the complex projective manifold of the 
full Hilbert space (i.e., $CP^N$, if $N$ is the Hilbert space
dimension).}) metric tensor -- which describes the absolute value of the overlap
amplitude between neighboring ground states \cite{Provost1980_1}.  
This metric plays an important role in understanding
the physics of quantum many-body ground states 
\cite{Thouless1998_1,MaUnpub2012_1} and is at the heart of current
research in quantum information theory \cite{Zanardi2007_1,You2007_1,Dey2012_1}.  For instance,
the diagonal components of the quantum metric tensor are none other than fidelity susceptibilities, whose scaling in the vicinity  of quantum phase transitions -- including topological phase transitions -- is
an object of great interest~\cite{Yang2008_1,Garnerone2009_1}.

The purpose of this 
paper is to understand the quantum geometry of a simple model,
the spin-1/2 XY chain in a transverse field.  For this integrable 
model, we solve the geometry and topology of the ground state manifold as
a function of three parameters: transverse magnetic field, interaction anisotropy,
and spin rotation about the transverse axis.  Using a standard trick from
Riemann geometry, we analyze the three-dimensional metric
by taking a series of two-dimensional cuts.  For cuts along which the Riemannian
manifold is regular, we identify the shape
of the manifold and show that the shape of each phase
is protected against symmetry-respecting perturbations 
by critical scaling theory of the metric tensor.  For
cuts along which the manifold is singular, we identify and classify the singularities.
As with the other cuts, we demonstrate that the singularities are 
robust against a variety of modifications to the low-energy theory.  We see three
types of geometric singularities: integrable, conical, and curvature singularities. 
Finally, we detail general circumstances where each type of singularity will arise.

Given the importance of the quantum metric tensor in understanding the
properties of ground state manifolds, it is surprising that there have
been no direct experimental measurements of the metric tensor to date.  Therefore, 
at the end of the paper we discuss several different proposals for experimentally measuring 
the components of the metric tensor. The first method is based on the direct representation 
of the metric tensor through the noise spectral function, generalizing a recent proposal
by Neupert et al.~\cite{NeupertArxiv2013_1} for measuring the metric tensor in non-interacting
Bloch bands.  The second method relates the metric tensor to a measurement of the leading  non-adiabatic contribution 
to the excess heat for square root ramps (see also  Refs. \citenum{DeGrandi2011_1} and 
\citenum{DeGrandiUnpub2013_1}). The third method similar identifies the metric with
leading non-adiabatic corrections to energy fluctuations for generic 
linear ramps.  Finally, the fourth method is based on analyzing the probability of doing zero 
work in single or double quenches, which is related to the time average of the 
well-known Loschmidt echo \cite{Silva2008_1}.  Using these techniques
the full many-body metric tensor is -- at least in principle -- experimentally accessible.
\footnote{It bears mentioning that there is a different definition of the metric tensor derived from
the Fisher information metric for density matrices. It can also be measured in practice, as it is related to standard susceptibilities
(cf. \rcite{Crooks2007_1}, Eq. (4)).  As such, the classical metric components clearly
are singular at phase transitions, as a result of critical scaling of the susceptibilities.}
We note that these measurement proposals do not rely on many of the geometric notions discussed elsewhere
in the paper, so those primarily interested in measuring the metric can skip directly 
\sref{sec:measuring_metric}.
We also note that the metric tensor can be readily extracted numerically as a non-adiabatic 
response of physical observables to imaginary time ramps~\cite{DeGrandi2011_1}, 
by directly evaluating overlaps of the ground state wave  functions at slightly different
couplings \cite{Venuti2007_1}, or through numerical integration of imaginary time noise
spectra of the generalized forces \cite{Schwandt2009_1,Albuquerque2010_1}.

The rest of this paper proceeds as follows.   
In \sref{sec:geom_def}, we introduce the metric tensor and show its relation to 
the Berry connection operators.  In \sref{sec:geom_invariants}, we describe how the Euler characteristic, 
curvature, and other geometric invariants are obtained from this metric.  
As a useful example, in \sref{sec:metric_xy} we explicitly solve for these
quantities for the case of the integrable quantum XY model in 
a transverse field. In \sref{sec:visualizing}, we show how to
visualize the metric manifold by mapping to an isometric surface
embedded in three dimensions.  These shapes motivate us to 
define invariant integrals consisting of the contribution to the
Euler characteristic within a given phase.  We
solve this exactly for the XY model then, in \sref{sec:universality},
we argue based on critical scaling of the metric tensor that the 
geometric integrals remain unchanged for all models in the same 
universality class.  To further understand the geometry of these
Riemann manifolds, we classify three types of singularity
in the Gaussian curvature that can occur in the vicinity of 
phase transitions: integrable (\sref{sec:singularities_integrable}), 
conical (\sref{sec:singularities_conical}),
and curvature (\sref{sec:singularities_curvature}) singularities.  Abstracting away from
the XY model, we detail situations under which each type of singularity should arise.
Finally, in \sref{sec:measuring_metric} we discuss different methods for measuring  
the quantum metric in terms of a more traditional condensed matter 
measurement of noise correlations, as well as through real-time 
ramps and quenches of the system parameters as is more relevant to 
isolated cold atom experiments.

\section{Geometry of the ground state manifold}
\label{sec:geom_def}

Consider a manifold of Hamiltonians described by some coupling
parameters $\vec\lambda$. A natural measure of the
distance between the ground state wave functions $|\psi_0 \ket$ 
separated by infinitesimal $d\vec\lambda$ is\cite{Provost1980_1}
\be
ds^2=1-|\bra
\psi_0(\vec\lambda)|\psi_0(\vec \lambda+d \vec \lambda)\ket|^2 =
\sum_{\mu \nu}\chi_{\mu\nu} d \lambda^\mu d \lambda^\nu,
\label{eq:chi1}
\ee 
where $\chi_{\mu \nu}$ is the geometric tensor:
\be
\chi_{\mu\nu}=\bra
\psi_0|\overleftarrow{\partial_\mu}\partial_\nu|\psi_0\ket-
\bra
\psi_0|\overleftarrow{\partial_\mu}|\psi_0\ket\bra\psi_0|\partial_\nu|\psi_0\ket ~,
\label{eq:chi}
\ee 
with $\partial_\mu \equiv \frac{\partial}{\partial \lambda^\mu}$.  As noted by
Provost and Vallee \cite{Provost1980_1}, this tensor is invariant under arbitrary 
$\lambda$-dependent $U(1)$ gauge transformation of the ground state wave functions.

Strictly speaking, \eref{eq:chi1} utilizes only the real symmetric
part of $\chi_{\mu\nu}$, which defines the metric tensor
associated with the ground state manifold: 
\be
g_{\mu\nu}=\Re[\chi_{\mu\nu}]={\chi_{\mu\nu}+\chi_{\nu\mu}\over
  2}.  
\ee 
However, in another seminal work~\cite{Berry1984_1}, Berry
introduced the notion of geometric phase (a.k.a. Berry phase) and the
related Berry curvature, which is given by the imaginary
(antisymmetric) part of the geometric tensor: 
\be
F_{\mu\nu}=-2\Im[\chi_{\mu\nu}]=i(\chi_{\mu\nu}-
\chi_{\nu\mu})=\partial_\mu A_\nu-\partial_\nu A_\mu~, 
\ee 
where $A_\mu=i\bra
\psi_0|\partial_\mu|\psi_0\ket$ is the Berry connection
within the ground state manifold. The Berry phase is just a
line integral of the Berry connection or -- by Stokes theorem -- a surface
integral of the Berry curvature: 
\be 
\Phi=\oint_{\partial S} \vec A \cdot
d\vec\lambda=\int_{S} F_{\mu\nu} dS_{\mu\nu}~, 
\ee 
where $dS_{\mu\nu}$ is a directed surface element.

In this work, we will be primarily interested in the metric tensor, $g_{\mu \nu}$.
One simple physical interpretation of the metric tensor is that it sets natural
units, allowing one to compare different physical parameters.
For example, if we consider the ground state manifold as a function of
magnetic field and pressure, one can ask how one Tesla compares to one Pascal. 
In the absence of a simple single particle coupling, the method for scaling these
quantities to compare them is not obvious.  However, the metric tensor provides
a natural answer by allowing one to compare the effects of these couplings on the
ground state fidelity.  Rescaling the units by the corresponding diagonal components 
of the metric tensor -- a.k.a. the fidelity susceptibilities -- one sets
natural units for different couplings.  Therefore, one Tesla can be compared to
one Pascal by comparing the "dimensionless"
couplings after rescaling $d\lambda_\mu\to d\lambda_\mu/\sqrt{g_{\mu\mu}}$.

\subsection{Geometric invariants of the metric tensor}
\label{sec:geom_invariants}

Underlying the classification of most topological phases is
the fact that the Berry curvature satisfies the Chern theorem \cite{Thouless1994_1}, which states that
the integral of the Berry curvature over a closed two-dimensional manifold
$\mathcal M$ in the parameter space is $2\pi$ times
an integer $n$, known as the Chern number: 
\be 
\oint_{\mathcal M} F_{\mu\nu} dS_{\mu\nu}=2\pi n~.  
\ee 
Physically this theorem reflects the single valuedness of the
wave function during adiabatic evolution: imagine splitting $\mathcal M$
into ``upper'' and ``lower''
surfaces. To maintain single valuedness, the Berry phases
obtained by integrating the Berry
curvature over the upper and lower surfaces can only be different by a
multiple of $2\pi$.

While the Berry curvature and its associated geometry are certainly of great interest
in modern condensed matter physics, the metric tensor $g_{\mu \nu}$
also plays an important role.  This tensor defines a Riemannian manifold
associated with the ground states, and it is interesting
to similarly inquire about its geometry and topology.
In particular, the shape of the Riemannian manifold
defines a different topological number, given by 
applying the Gauss-Bonnet theorem\cite{Carmo1976_1}
to the quantum metric tensor:
\be
\frac{1}{2\pi} \left [\int_{\mathcal M} K dS+\oint_{\partial \mathcal M} k_g
dl \right]=\chi(\mathcal M)~,
\label{eq:gauss_bonnet}
\ee 
where $\chi(\mathcal M)$ is the integer Euler characteristic
describing the topology of the manifold $\mathcal M$ with metric
$g_{\mu \nu}$.  The two terms on the left side of \eref{eq:gauss_bonnet}
are the bulk and boundary contributions to the Euler characteristic
of the manifold.  We refer to the first term, 
\be
\chi_\mathrm{bulk}(\mathcal M) = \frac{1}{2\pi} \int_{\mathcal M} K dS ~,
\label{eq:chi_bulk}
\ee
and the second term,
\be
\chi_\mathrm{boundary}(\mathcal M) = \frac{1}{2\pi} \oint_{\partial \mathcal M} k_g dl ~,
\label{eq:chi_boundary}
\ee
as the bulk and boundary Euler integrals, respectively.  These terms, 
along with their constituents -- the Gaussian curvature ($K$), the geodesic curvature ($k_g$), the area
element ($dS$), and the line element ($dl$) -- are \emph{geometric invariants}, meaning that
they remain unmodified under any change of variables.  More explicitly,
if the metric is written in first fundamental form as
\be
ds^2 = E d\lambda_1^2 + 2 F d\lambda_1 d\lambda_2 + G d\lambda_2^2 ~,
\ee
then these invariants are given by \cite{Kreyszig1959_1}
\beq
\nn
K & = & \frac{1}{\sqrt g} \left[ \frac{\partial}{\partial \lambda_2} \left( 
\frac{\sqrt g \, \Gamma^2_{11}}{E} \right) - \frac{\partial}{\partial \lambda_1} \left(
\frac{\sqrt g \, \Gamma^2_{12}}{E} \right) \right] \\
\nn
k_g &=& \sqrt{g} G^{-3/2} \Gamma^1_{22} \\
\nn
dS &=& \sqrt g d\lambda_1 d\lambda_2 \\
dl &=& \sqrt G d\lambda_2 ~,
\label{eq:invariants}
\eeq
where $k_g$ and $dl$ are given for a curve of constant $\lambda_1$.  The metric 
determinant $g$ and Christoffel symbols $\Gamma^k_{ij}$ are 
\beq
g&=&EG-F^2 \\
\Gamma^{k}_{ij} &=& \frac{1}{2} g^{km} \left( \partial_j g_{im}
+ \partial_i g_{jm} - \partial_m g_{ij} \right) ~,
\eeq
where $g^{ij}$ is the inverse of the metric tensor $g_{ij}$.  

As we will see, in general the bulk and the 
boundary terms are not individually protected against perturbations
for an arbitrary manifold $\mathcal M$.  
A major purpose of the current work is to demonstrate that if the parameter 
space manifold terminates at a phase boundary, however, then not only is the sum 
in \eref{eq:gauss_bonnet} protected against various perturbations,
but so is each term individually. Thus, for example, the bulk Euler integral
(\eref{eq:chi_bulk}) can be used for classification of geometric properties of 
different phases. This geometric invariant is in general different from the 
Chern number, and can be non-trivial even in the absence of time reversal 
symmetry breaking.  

While the Gauss-Bonnet theorem has a higher dimensional generalization known as the 
Chern-Gauss-Bonnet theorem \cite{Chern1945_1}, in this work we will focus 
only on the two-dimensional version. We emphasize that the dimensionality 
here is that of parameter space; the physical dimensionality of the system 
can be arbitrary. 
The choice of parameters is also arbitrary, and is usually dictated 
either by experimental accessibility, symmetry properties of the system, or 
other related considerations. Choosing appropriate parameters for studying 
geometric properties of the phases is therefore similar to choosing 
parameters defining the phase diagram.

\subsection{Defining the geometric tensor through gauge potentials}
\label{sec:gauge_Pot}

It can be convenient to express the geometric tensor through the  
Berry connection operators $\mathcal A_\mu=i\partial_\mu$ associated with the couplings 
$\lambda_\mu$, which one can think of as gauge potentials in parameter space.
 These gauge operators are formally defined through the matrix elements
\be
\mathcal A_\mu^{mn}=i \bra m| \partial_\mu |n\ket ~,
\label{guage_pot}
\ee
which implicitly depend on the $U(1)$ phase choice for each energy
eigenstate $|n\ket$ at each $\vec \lambda$.  
If the basis dependence of $\vec\lambda$ is expressed through a unitary rotation of 
some parameter-independent basis,
\[
|n(\vec\lambda)\ket = U^{nm}(\vec\lambda) |m\ket_0,
\]
then the gauge potentials can be written as
\be
\mathcal A_\mu=i U^\dagger \partial_\mu U.
\label{gauge_pot1}
\ee

The operator $\mathcal A_\mu$ generates infinitesimal translations of the basis 
vectors within the parameter space. For instance, if spatial coordinates play 
the role of parameters, then the corresponding gauge potential is the momentum operator.
If the parameters characterize rotational angles, the gauge potential is the angular momentum operator. 

The ground state expectation value of the gauge operator is by definition the Berry connection
\be
A_\mu=\bra \psi_0| \mathcal A_\mu|\psi_0\ket.
\ee
The geometric tensor is the expectation value of their covariance matrix:
\beq
\chi_{\mu\nu}&=&\bra \psi_0| \mathcal A_\mu \mathcal A_\nu|\psi_0\ket_c\nonumber \\ 
&\equiv&\bra \psi_0| \mathcal A_\mu \mathcal A_\nu|\psi_0\ket-
\bra \psi_0| \mathcal A_\mu|\psi_0\ket\bra \psi_0| \mathcal A_\nu|\psi_0\ket.
\eeq
More explicitly, the components of the metric tensor and the Berry curvature 
are expressed through the connected expectation value of the anti-commutator 
and the commutator of the gauge potentials, respectively:
\beq
&&g_{\mu\nu}={1\over 2} \bra \psi_0|\mathcal A_\mu\mathcal A_\nu+\mathcal A_\nu\mathcal A_\mu|\psi_0\ket_c,\\
&& F_{\mu\nu}=i\bra \psi_0|\mathcal A_\mu\mathcal A_\nu-\mathcal A_\nu\mathcal A_\mu|\psi_0\ket.
\eeq

Although we will be interested only in the ground state manifold for
the remainder of this paper, we briefly comment that the definitions above can 
be extended to arbitrary stationary or non-stationary density matrices.  For
example, with a finite temperature equilibrium ensemble, one can define
$F_{\mu\nu}=i{\rm Tr} (\rho_\mathrm{thermal} [\mathcal A_\mu,\mathcal A_\nu])$. 

In \sref{sec:metric_xy}, we will explicitly calculate the gauge potentials of the 
quantum XY chain. Here we comment on a few of their general properties. 
First, we note that gauge potentials are Hermitian operators. This follows from differentiating 
the identity $\bra n(\vec \lambda)|m(\vec \lambda)\ket=\delta_{nm}$ with respect to $\lambda_\mu$, or more directly  from their definition in terms of unitaries (\eref{gauge_pot1}). 
Also, the gauge potentials satisfy requirements of locality. In particular, 
if the Hamiltonian can be written as the sum of local terms $\mathcal H=\sum_i  h_i$ and $\vec \lambda$ 
represent global couplings within this Hamiltonian, then the geometric tensor is extensive
\footnote{We note that the proof of extensivity in \rcite{Venuti2007_1} applies to arbitrary 
finite temperature states as long as the corresponding expectation values of the correlation 
functions $\bra d_\lambda h_i(t) d_\mu h_j(0)\ket_c$ decay sufficiently fast with respect 
to both time $t$ and spatial separation $|i-j|$. This is usually the case except near critical points.}.  
In particular, this implies that fluctuations of the gauge potentials are also extensive, 
which is a general property of local extensive operators.
Similarly, if $\vec\lambda$ represent local (in space) perturbations,  then the geometric tensor is generally system size independent, so that $\mathcal A_\mu$ is again local. As usual, various singularities -- including those breaking locality
-- can develop in gauge potentials near phase transitions. Finally, we point out that if 
$\lambda_\mu$ is a symmetry of the Hamiltonian, i.e., if the Hamiltonian is invariant under 
$\lambda_\mu \to \lambda_\mu+\delta\lambda_\mu$, then all gauge potentials $\mathcal A_\nu$ 
are also invariant under this symmetry.

\subsection{Metric tensor of the quantum XY chain}
\label{sec:metric_xy}

As our primary example, 
we consider a quantum XY chain described by the
Hamiltonian
\be 
\mathcal H=-\sum_j \big[ J_x \mathrm
s_{j}^x s_{j+1}^x + J_y \mathrm
s_{j}^y s_{j+1}^y + h s_{j}^z \big] ~,
\ee 
where $J_{x,y}$ are exchange couplings, $h$ is a transverse field, and the 
spins are represented as Pauli matrices $s^{x,y,z}$.
It is convenient to re-parameterize the model in terms of new couplings $J$
and $\gamma$ as
\be 
J_x=J \left( \frac{1+\gamma}{2} \right),\; J_y=J \left( \frac{1-\gamma}{2} \right)~,
\ee 
where $J$ is the energy scale of the exchange interaction and $\gamma$ is 
its anisotropy. We add an additional tuning parameter $\phi$, corresponding
to simultaneous rotation of all the spins about the $z$-axis by
angle $\phi/2$.
While rotating the angle $\phi$ has no effect on the 
spectrum of $H$, it does modify the ground state wave function.
To fix the overall energy scale, we set $J=1$. 

\begin{figure}
\includegraphics[width=.6\linewidth]{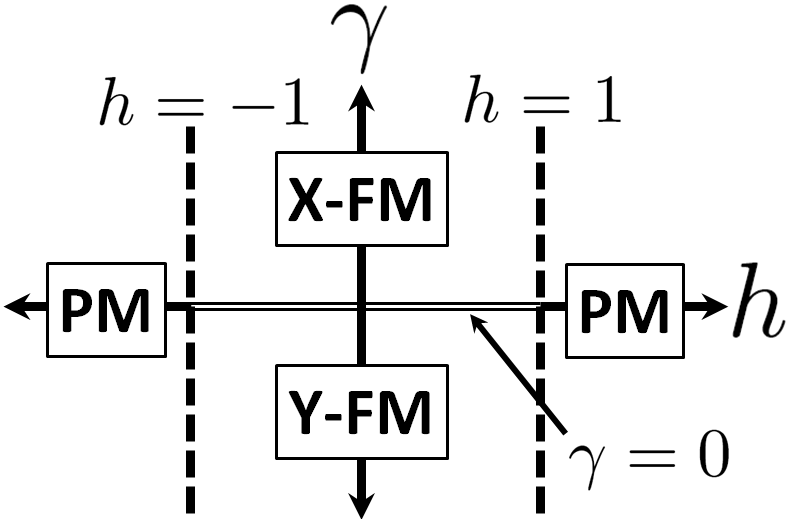}
\caption{Ground state phase diagram of the XY Hamiltonian (\eref{eq:H_XY})
for $\phi=0$.  The rotation parameter $\phi$ modifies the Ising 
ferromagnetic directions, otherwise maintaining all features of the phase diagram.  As a function
of transverse field $h$ and anisotropy $\gamma$, the ground state undergoes
continuous Ising-like phase transitions between paramagnet and ferromagnet at $h=\pm 1$
and anisotropic transitions between ferromagnets aligned along 
X and Y directions (X/Y-FM) at $\gamma=0$.
These two types of phase transition meet at multi-critical points, which are 
described in detail in \rcite{Mukherjee2011_1}.}
\label{fig:phasediagram}
\end{figure}

The Hamiltonian described above can be written as
\beq
 &&\mathcal H(h,\gamma,\phi)=-\sum_j \left[
s_{j}^+s_{j+1}^-+{\rm h.c.} \right] \nonumber\\ 
&&-\gamma \sum_j
\left[ e^{i\phi} s_j^+s_{j+1}^+ + {\rm h.c.} \right] - h\sum_j
s_{j}^z ~.
\label{eq:H_XY}
\eeq
Since the Hamiltonian is invariant under the 
mapping $\gamma \to -\gamma$, $\phi \to \phi + \pi$, we generally
restrict ourselves to $\gamma \geq 0$, although we occasionally plot the
superfluous $\gamma < 0$ region when convenient. 
This model has a rich phase diagram~\cite{Damle1996_1, Mukherjee2011_1},
as shown in \fref{fig:phasediagram}. There is a
phase transition between paramagnet and Ising ferromagnet at $|h|=1$ and
$\gamma \neq 0$. There is an additional critical line at the
isotropic point ($\gamma=0$)
for $|h|<1$. The two transitions meet at multi-critical points when $\gamma=0$
and $|h|=1$. Another notable line is $\gamma=1$, 
which corresponds to the transverse-field Ising (TFI) chain. 
Finally let us note that there are two other special lines $\gamma=0$ and $|h|>1$ 
where the ground state is fully polarized along the magnetic field and thus $h$-independent. 
Thus this line is characterized by vanishing susceptibilities including vanishing metric 
along the $h$-direction. As we discuss in \sref{sec:K_symm_line} such state is fully 
protected by the rotational symmetry of the model and can be terminated only at the critical (gapless) point.
The phase diagram is invariant under changes of the rotation
angle $\phi$. 

Rewriting the spin Hamiltonian in terms of free fermions via a
Jordan-Wigner transformation, $\mathcal H$ can be mapped to an 
effective non-interacting spin one-half model\cite{Sachdev1999_1} with 
\beq 
&&\mathcal
H=\sum_k \mathcal H_k \label{eq:hk_xxz} ~ ; \\ 
&&\mathcal H_k=-\left(
\begin{array}{cc}
h-\cos(k) & \gamma \sin(k) \mathrm e^{i\phi}\\ 
\gamma \sin(k)
\mathrm e^{-i\phi} & -[h-\cos(k)]
\end{array}
\right)\nonumber 
\eeq
This mapping yields a unique ground state throughout the phase diagram by 
working in a particular fermion parity sector\cite{Lieb1961_1}; none of the conclusions
below will change if the other sector is chosen in cases when the ground
state is degenerate.  A more general analysis involving the non-Abelian 
metric tensor\cite{Ma2010_1,NeupertArxiv2013_1} 
is outside the scope of this work.

The ground state of $\mathcal H_k$ is a Bloch vector with azimuthal angle $\phi$
and polar angle 
\be
\theta_k = \tan^{-1} \left[ \frac{\gamma \sin(k)}{h-\cos(k)} \right ]~.
\ee
To derive the components of the metric tensor, we start by considering the 
gauge operators introduced in \sref{sec:gauge_Pot}: $\mathcal A_\mu \equiv i \partial_\mu$.
If we consider the transverse field $h$, 
we see that
\be
\partial_h |\mathrm{gs}_k\ket = \frac{\partial_h \theta_k}{2} 
\left( \begin{array}{c} -\sin\left( \frac{\theta_k}{2}\right) e^{i\phi/2} \\
\cos \left( \frac{\theta_k}{2}\right) e^{-i\phi/2} \end{array} \right) = 
-\frac{\partial_h \theta_k}{2} |\mathrm{es}_k\ket ~.
\ee
The same derivation applies to the anisotropy $\gamma$, since 
changing either $\gamma$ or $h$ only modifies
$\theta_k$ and not $\phi$. Thus we find
\be
\mathcal A_\lambda = \frac{1}{2} \sum_k \big( \partial_\lambda \theta_k \big) \tau_k^y ~,
\ee
where $\lambda=\{h,\gamma\}$ and $\tau_k^{x,y,z}$ are Pauli matrices that act in the 
instantaneous ground/excited state basis, i.e., $\tau_k^z |\mathrm{gs}_k\ket=
|\mathrm{gs}_k\ket$,  $\tau^z_k |\mathrm{es}_k\ket=-|\mathrm{es}_k\ket$. 
Similarly, for the parameter $\phi$, we find that
\be
\mathcal A_\phi = -\frac{1}{2} \sum_k \left[ \cos(\theta_k) \tau^z_k + \sin(\theta_k) \tau^x_k \right].
\ee

In terms of these gauge potentials, the metric tensor and Berry curvature
can be written as (see \sref{sec:gauge_Pot})
\be
g_{\mu \nu} = \frac{1}{2} \bra \{\mathcal A_\mu, \mathcal A_\nu \} \ket_c ~~,~~
F_{\mu \nu} = i \bra [\mathcal A_\mu,\mathcal A_\nu ] \ket ~.
\ee
In the case of the XY model, the metric tensor reduces to
\beq 
&& g_{hh}={1\over 4}\sum_k
\left({\partial\theta_k\over\partial
  h}\right)^2 ~; ~~~
 g_{\gamma \gamma}={1\over
  4}\sum_k \left({\partial\theta_k\over\partial
  \gamma}\right)^2\nonumber\\ 
&& g_{h\gamma}={1\over
  4}\sum_k {\partial\theta_k\over\partial h}{\partial\theta_k\over
  \partial \gamma}\nonumber
~; ~~~ g_{\phi\phi}={1\over 4}\sum_k
     \sin^2(\theta_k)\nonumber\\ 
&& g_{h\phi}= g_{\gamma \phi}=0 
\eeq 
Although we will not be interested in the Berry curvature in this work, 
we show the corresponding expressions for completeness
\beq
&&F_{h\phi}=-{1\over 2}\sum_k {\partial\theta_k\over \partial h}\sin(\theta_k),\nonumber\\
&&F_{\gamma\phi}=-{1\over 2}\sum_k {\partial\theta_k\over \partial \gamma}\sin(\theta_k),\;
F_{h \gamma}=0.
\eeq

The expressions for the metric tensor can be evaluated
in the thermodynamic limit, where the summation becomes
integration over momentum space. It is convenient to
divide all components of the metric tensor by the system size
and deal with intensive quantities $g_{\mu \nu} \to 
 g_{\mu \nu} / L$. Then one calculates these integrals to find that
\beq 
&& g_{\phi\phi}= \frac{1}{8} \left\{
\begin{array}{cc}
\frac{|\gamma|}{|\gamma|+1}, &
|h|<1\\ \frac{\gamma^2}{1-\gamma^2}\left( \frac{|h|}
{\sqrt{h^2-1+\gamma^2}} - 1 \right) , &
|h|>1
\end{array}
\right.
\label{eq:g_phiphi}\nonumber\\
&&  g_{hh}= \frac{1}{16} \left\{\begin{array}{cc}
      \frac{1}{|\gamma| (1-h^2)}, & |h|<1\\ \frac{|h| \gamma^2}
{(h^2-1)(h^2-1+\gamma^2)^{3/2}}, & |h|>1
\end{array}\right. \label{eq:g_hh}\nonumber\\
&& g_{\gamma \gamma}= \frac{1}{16} \left\{\begin{array}{cc}
      \frac{1}{ |\gamma| (1+|\gamma|)^2}, & |h|<1\\ \left( \begin{array}{cc} 
	\frac{2}{(1-\gamma^2)^2} \Big[ \frac{|h|}{\sqrt{h^2-1+\gamma^2}} - 1 \Big] - \\
	 \frac{|h| \gamma^2}{(1-\gamma^2)(h^2-1+\gamma^2)^{3/2}}
\end{array} \right), & |h|>1
\end{array}\right. \label{eq:g_gg}\nonumber\\
&&  g_{h\gamma}= \frac{1}{16} \left\{\begin{array}{cc} 0, &
|h|<1\\ \frac{-|h| \gamma}{h (h^2-1+\gamma^2)^{3/2}} , & |h|>1
\end{array}\right. \label{eq:g_hg}
\eeq


\begin{figure}
\includegraphics[width=.7\linewidth]{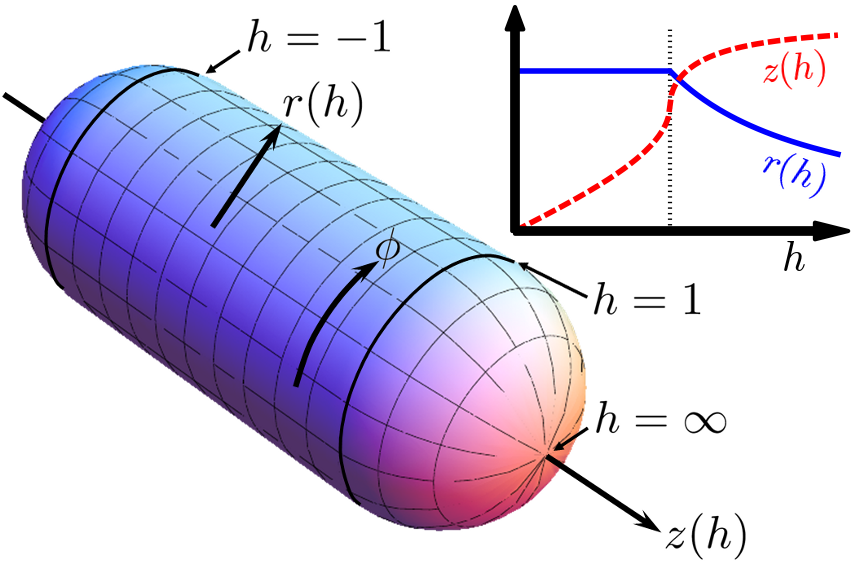}
\caption{Equivalent graphical representation of the phase diagram of
  the transverse field Ising model ($\gamma=1$) in the $h-\phi$ plane (\eref{eq:XY_rz}). 
  The ordered ferromagnetic phase maps to a cylinder of constant radius.
  The   disordered paramagnetic phases $h>1$ and $h<-1$ map to 
  the two hemispherical caps. The inset shows how the cylindrical
  coordinates $z$ and $r$ depend on the transverse field $h$.}
\label{fig:ising_h_phi}
\end{figure}

\subsection{Visualizing the ground state manifold}
\label{sec:visualizing}

Using the metric tensor we can visualize the ground
state manifold by building an equivalent (i.e., isometric) surface
and plotting its shape. It is convenient to
focus on a two-dimensional manifold by fixing
one of the parameters. We then represent the two-dimensional 
manifold as an equivalent three-dimensional surface.
To start, let's fix the anisotropy parameter
$\gamma$ and consider the $h-\phi$ manifold. Since the metric tensor has cylindrical
symmetry, so does the equivalent surface. Parameterizing our shape
in cylindrical coordinates and requiring that
\be
dz^2+dr^2+r^2d\phi^2=g_{hh}dh^2+g_{\phi\phi} d\phi^2 ~,
\ee 
we see that
\be 
r(h)=\sqrt{g_{\phi\phi}},\;z(h)=\int_0^h dh_1
\sqrt{g_{hh}(h_1)-\left({dr(h_1)\over dh_1}\right)^2}.  
\ee 
Using \eref{eq:g_phiphi}, we explicitly find
the shape representing the XY chain. In the Ising limit ($\gamma=1$), we get
\beq 
&&\left\{
\begin{array}{l}
r(h)={1\over 4}\\ z(h)={\arcsin(h) \over 4}
\end{array}
\right.\quad |h|<1,\nonumber\\ 
&& \left\{
\begin{array}{l}
r(h)={1\over 4 |h|}\\ z(h)={\pi\over 8}{|h|\over h}+{\sqrt{h^2-1}\over
  4 h}
\end{array}
\right.\quad |h|>1  ~.
\label{eq:XY_rz}
\eeq 
The phase diagram is thus represented by a
cylinder of radius $1/4$ corresponding to the ferromagnetic phase
capped by the two hemispheres representing the
paramagnetic phase, as shown in \fref{fig:ising_h_phi}. It is easy to check
that the shape of each phase does not depend on the anisotropy
parameter $\gamma$, which simply changes the aspect ratio and radius of the
cylinder. Because of the relation $r(h)=\sqrt{g_{\phi\phi}}$ this 
radius vanishes as the anisotropy parameter $\gamma$ goes to zero.
By an elementary integration of the Gaussian curvature,
the phases have bulk Euler integral $0$ for the ferromagnetic cylinder
and $1$ for each paramagnetic hemisphere.  These numbers add up to $2$
as required, since the full phase diagram is homeomorphic
to a sphere. From~\fref{fig:ising_h_phi},
it is also clear that the phase boundaries at $h=\pm 1$ are 
geodesics, meaning that the geodesic curvature (and thus 
the boundary contribution $\chi_\mathrm{boundary}$) is zero for a
contour along the phase boundary.  As we will soon see, this boundary
integral protects the value of the bulk integral and vice versa.

In the Ising limit ($\gamma=1$), the 
shape shown in~\fref{fig:ising_h_phi}, can also be easily seen from computing the curvature $K$
using \eref{eq:invariants}.  Within the ferromagnetic phase, the 
curvature is zero -- no surprise, given that the metric is flat
by inspection.  The only shape with zero curvature and
cylindrical symmetry is a cylinder.  Similarly, within the 
paramagnet, the curvature is a constant $K=16$, like that
of a sphere.  Therefore, to get cylindrical symmetry, the 
phase diagram is clearly seen to be a cylinder capped by two hemispheres.

\begin{figure}
\includegraphics[width=\linewidth]{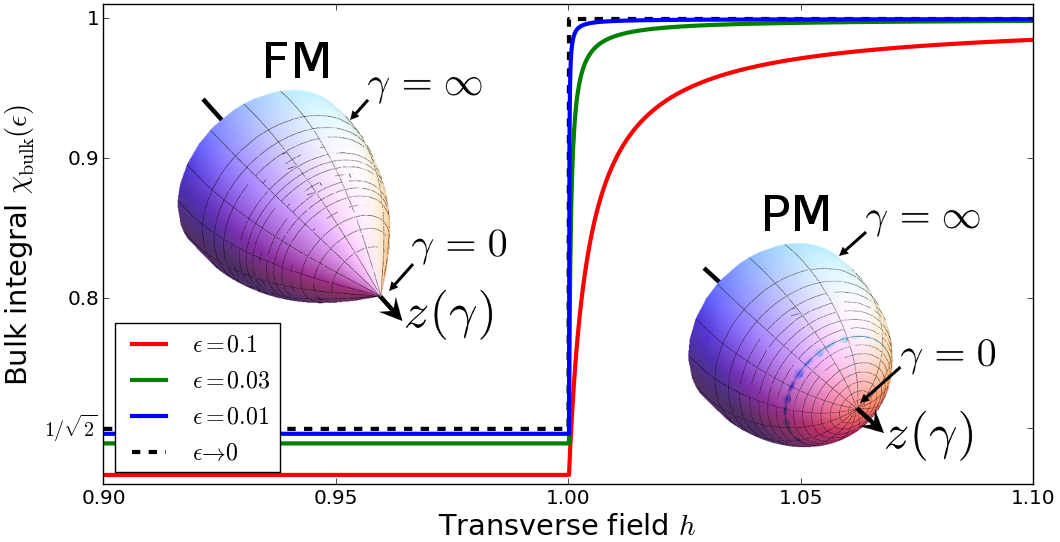}
\caption{(insets) Equivalent graphical representation of the phase diagram of
  the XY model in the $\gamma-\phi$ plane, where $\gamma\in[0,\infty)$ and
  $\phi\in[0,2\pi]$. The right inset shows the paramagnetic
  disordered phase and the left inset represents the
  ferromagnetic phase. It is clear that in the latter case
  there is a conical singularity developing at $\gamma=0$ which represents
  the anisotropic phase transition.
  The plots show bulk Euler integral $\chi_\mathrm{bulk}(\epsilon)$
  as defined in \eref{eq:euler_bulk},
  demonstrating the jump in $\chi_\mathrm{bulk}$ at the phase transition
  between the paramagnet and ferromagnet in the limit $\epsilon \to 0^+$.}
\label{fig:anising_g_phi}
\end{figure}

We can also reconstruct an equivalent shape in the $\gamma-\phi$ plane. In
this case we expect to see a qualitative difference for $|h|>1$ and
$|h|<1$ because in the latter case there is an anisotropic phase
transition at the isotropic point $\gamma=0$, while in the former case
there is none. These two shapes are shown in
\fref{fig:anising_g_phi}. The anisotropic phase transition is manifest
in the conical singularity developing at $\gamma=0$.\footnote{
We note a potential point of confusion, namely that a naive application of
\eref{eq:invariants} would seem to indicate that the curvature is a 
constant $K=4$ in the ferromagnetic phase for $\gamma>0$, in which case the singularity at
$\gamma=0$ is not apparent.  However, a more careful derivation shows that
the curvature is indeed singular at $\gamma=0$: 
$K=4-8(1-\gamma)\frac{\partial^2}{\partial^2 \gamma} |\gamma|
=4-16\delta(\gamma)$, where $\delta(\gamma)$ is the Dirac delta function.}

The singularity at $\gamma=0$ yields a non-trivial bulk Euler integral
for the anisotropic phase transition. To see this, consider the bulk integral
\be 
\chi_\mathrm{bulk}(\epsilon)=\lim_{L\to\infty}\int_0^{2\pi} d\phi
\int_{\epsilon}^{\infty} d \gamma \, \sqrt{g(\gamma,\phi)}K(\gamma,\phi) ~.
\label{eq:euler_bulk}
\ee 
In the limit $\epsilon \to 0^+$,
this integral has a discontinuity as a function of $h$ at the phase transition, 
as seen in \fref{fig:anising_g_phi}. Thus, $\chi_\mathrm{bulk}
\equiv \chi_\mathrm{bulk}(\epsilon=0^+)$ can be used as a geometric
characteristic of the anisotropic phase transition. As we will show in the next section, its value is 
$\chi_\mathrm{bulk}=1/\sqrt 2$ in the ferromagnetic phase and $\chi_\mathrm{bulk}=1$ in the 
paramagnetic phase. This non-integer geometric invariant is due to the existence of 
a conical singularity.

The last two-dimensional cut, namely the $h-\gamma$ plane at fixed $\phi$,
is significantly more complicated and we have not been able to find any simple
shape to represent this part of the phase diagram. However, using the technology
that we develop below for the more easily visualized surfaces, we analyze the 
$h-\gamma$ plane in \sref{sec:singularities_curvature}.

\section{Universality of the Euler integrals}
\label{sec:universality}

We now wish to show that the Euler integrals characterizing
various phases of the XY model are universal to such phase transitions
due to critical scaling of the metric.
We begin by considering the transverse-field
Ising (TFI) model with $\gamma=1$, $h\in(-\infty,\infty)$, and $\phi\in[0,2\pi)$.
For this model, it is known\cite{Venuti2007_1} that the metric tensor, and thus the
associated curvatures, obey certain scaling laws near the QCP. Therefore,
since the boundary of the phase is at such a QCP, 
critical scaling theory is encoded in the boundary Euler integral.

However, knowing the boundary Euler integral is sufficient
to determine the bulk integral. To see this,
consider the region $h\in(-1+\epsilon,1-\epsilon)$
for small positive $\epsilon$. Since the region only spans a single phase,
there are no ground state degeneracies within this region, meaning
the $h-\phi$ surface is homeomorphic to an open cylinder. Because an open cylinder
has Euler characteristic $0$, the Gauss-Bonnet theorem becomes
\be
\chi_\mathrm{bulk}=-\sum_\mathrm{boundaries} \chi_\mathrm{boundary}.
\label{eq:bulk_bound_corr}
\ee
We want to solve for the bulk Euler integral,
in the limit that the boundaries of the region are taken to the phase
boundary ($\epsilon \to 0^+$). However,
according to \eref{eq:bulk_bound_corr}, the bulk Euler integral is just
minus the sum of the boundary integral, which
are much easier to solve for.  This ``bulk-boundary correspondence'' is
what allows us to use critical scaling theory to determine the bulk
Euler integral for each phase.

\subsection{Example: Exact metric of the XY chain}
\label{sec:euler_xy_exact}

As an initial demonstration of this method, consider the exact expression
for the metric of the TFI model, given in \eref{eq:g_hh}.
For a diagonal metric along a curve of constant $h$, the geodesic
curvature reduces to
\be
k_{g}^{h\mathrm{\, const.}}=\frac{\partial\sqrt{g_{\phi \phi}}/\partial h}{\sqrt{g_{\phi \phi} g_{hh} }}~.
\label{eq:liouville}
\ee
For the case $|h|>1$, this gives
\be
k_{g}(h)=\frac{-4/h^{2}}{\sqrt{\frac{1}{ h^{4}(h^{2}-1)}}}=
-4(h^{2}-1)^{1/2}\stackrel{|h|\to1}{\longrightarrow}0.
\label{eq:tfi_pm_geodesic}
\ee
Integrating over one of the critical lines, $h=\pm 1$ 
and $\phi\in [0,2 \pi)$, gives
$\chi_\mathrm{boundary} (|h|=1) = 0$.
To get some intuition as to what the boundary Euler integral of
zero means, consider the three-dimensional embedding shown in \fref{fig:ising_h_phi}.
A curve with $k_g=0$ is, by definition, a geodesic.  This makes sense, since
the circle at $h=1$ is clearly a geodesic of both the cylinder and the hemisphere.
In general, a smooth curve on a cylindrically-symmetric 
surface will be a geodesic if the radius is at a local extremum,
i.e., $dr/dz=0$.  This is clearly satisfied in the case of the TFI model, because 
$dr/dh$ is finite near the QCP, while $dz/dh\to\infty$ (see \fref{fig:ising_h_phi}, inset).

Similarly, for the limiting point at $h=\infty$, 
we can calculate the boundary Euler integral:
\beq
\chi_\mathrm{boundary}(h) &=& \frac{1}{2\pi}\oint k_{g}dl=
\frac{1}{2 \pi}\int_{0}^{2\pi}k_{g}(h) \sqrt{g_{\phi \phi}(h)}\; d\phi \nn \\
&=&-\sqrt{\frac{h^{2}-1}{h^{2}}}\stackrel{h\to\infty}{\longrightarrow}-1.
\eeq
By the same logic, $\chi_\mathrm{boundary}\to -1$ in the limit
$h\to-\infty$.  Therefore, using \eref{eq:bulk_bound_corr}, we quickly
obtain the $1-0-1$ breakup of the bulk Euler integral for the TFI model.

To further illustrate the analytical power of this method, we can now
compute the bulk Euler integral of the
$\gamma-\phi$ plane for arbitrary $|h|<1$, as defined in \eref{eq:euler_bulk}.
By a similar analysis as before, one finds that
\beq
\chi_{\mathrm{boundary}}(\gamma) & = &
\frac{1}{2 \pi}\int_{0}^{2\pi} k_{g}(\gamma)\sqrt{g_{\phi \phi}(\gamma)}\; d\phi \nn \\
&=&-\frac{1}{\sqrt{2(\gamma+1)}}\stackrel{\gamma \to 0}{\longrightarrow}-\frac{1}{\sqrt{2}} ~.
\eeq
Since the limit $\gamma \to \infty$
corresponds to a geodesic (see \fref{fig:anising_g_phi}), 
giving $\chi_\mathrm{boundary}(\gamma\to\infty) = 0$,
we see that the bulk Euler integral is 
$\chi_{\mathrm{bulk}}=\frac{1}{\sqrt{2}}$.

\subsection{Universality from critical scaling of the metric}
\label{sec:universality_critical_scaling}

We now use critical
scaling theory to find the Euler integrals of these phase
boundaries for more general models. Consider first 
the case of an arbitrary model in the TFI
(a.k.a. 2D Ising) universality class.  We know from the scaling theory
of \rcite{Venuti2007_1} that the metric diverges near the QCP
with a power law set by scaling dimension of the perturbing
operators, $\partial_{\lambda} H$. For example, in the transverse-field
direction, the metric must scale as $g_{hh}\sim|h-h_{c}|^{-1}$
for arbitrary models in the Ising universality class. Similarly, since
the parameter $\phi$ is marginal near the Ising critical point, the
singular part of $g_{\phi \phi}$ has scaling dimension $+1$.  Adding in
the regular part of $g_{\phi \phi}$ to get a non-zero value near the phase
transition, we see that to leading order $g_{\phi\phi}\sim A+B|h-h_{c}|$,
where $A$ and $B$ are constants. Plugging this into the formula for the 
boundary Euler integral, one finds that
\be
\chi_{\mathrm{boundary}}\sim\frac{\mathrm{const.}}{\sqrt{|h-h_{c}|^{-1}}}
\sim|h-h_{c}|^{1/2} \stackrel{h\to h_c}{\longrightarrow} 0.
\ee
Therefore, the boundary (and thus bulk) Euler integral of 
the Ising phase transition is protected
by the critical scaling properties of the metric tensor.
In terms of the geometry of the three-dimensional embedding,
adding irrelevant perturbations to the Hamiltonian will shift
the critical point and deform the shape away from the critical point.
However, the phase boundary between the ferromagnet and paramagnet
will remain a geodesic ($dr/dz$ will remain zero).  The fact that the geodesic curvature 
is zero on the phase boundary has an intuitive physical interpretation.  Consider two 
points on the phase boundary. The geodesic defines the line of the shortest distance 
between these two points in the Riemannian manifold defined by the metric tensor $g$. It 
is clear that this line should be entirely confined to the phase boundary 
since any deviations from it result in moving toward the direction of the relevant coupling, 
along which the metric tensor diverges. Since the phase boundary coincides with the geodesic, 
the geodesic curvature is zero by definition.

To understand the more complicated anisotropic direction, 
we expand the Hamiltonian around $\gamma=0$.
Close to the QCP ($|\gamma| \ll 1$), the spectrum is
gapless at a single momentum $k_0 = \cos^{-1}(h) \in (0,\pi)$,
around which we can linearize the equations.  Then the linearized mode 
Hamiltonian is 
\beq
\label{eq:Hkprime_anisotropic}
H_{k} &=& - \left( \begin{array}{cc} 
(k-k_0) \sin k_0 & \gamma e^{i \phi} \sin k_0 \\
\gamma e^{-i \phi} \sin k_0 & -(k-k_0) \sin k_0
\end{array}\right) \\ 
&=& -\left( (k-k_0) \sin k_0 \right) \sigma^z_{k} -
\left( \gamma \sin k_0 \right) 
\left[ \sigma^x_{k} \cos \phi  - \sigma^y_{k} \sin \phi \right] \nn
\eeq
where $\sigma^{(x,y,z)}$ are pseudo-spin Pauli matrices.
The presence of $\sin(k_0)$ in both terms suggests
fine-tuning, but this turns out to be unnecessary.  Therefore, we consider
a more general Hamiltonian of this form:
\be
H_{k'} = \beta k' \sigma_{k'}^z + 
\alpha \beta \gamma
\left[ \sigma^x_{k'} \cos \phi  + \sigma^y_{k'} \sin \phi \right] ~,
\label{eq:Hk_general}
\ee
where $\alpha$ and $\beta$ are arbitrary constants and $k'\equiv k-k_0$ is 
the momentum difference from the gapless point.  This linearized Hamiltonian has
\be
\theta_{k'} = \tan^{-1} \left( \frac{\alpha \gamma}{k'} \right)~.
\ee

The scaling limits of 
$g_{\gamma \gamma}$ and $g_{\phi \phi}$ are now relatively
straightforward to compute.  The formulas are, as before,
\be
 g_{\gamma \gamma} = \frac{1}{4 L} \sum_{k'} \left( 
\frac{\partial \theta_{k'}}{\partial \gamma} \right)^2 ~, \hspace{0.25in}
 g_{\phi \phi} = \frac{1}{4 L} \sum_{k'} \sin^2 (\theta_{k'}) ~.
\ee
In the thermodynamic limit, we convert the sum to an integral and
define the scaling variable
\be
\kappa \equiv k' / \gamma ~,
\ee
which goes from $\kappa=-\infty$ to $\infty$ in the scaling limit, $|\gamma| \ll 1$.
Thus,
\beq
g_{\phi \phi} & = & \frac{1}{8 \pi} 
\int_{k'} \sin^2 \theta_{k'} dk'
 = \int_{k'} \frac{\alpha^2 \gamma^2}{\alpha^2 \gamma^2 + k'^2} dk' \nn \\
& = & \frac{\alpha^2 \gamma}{8 \pi} \int_{-\infty}^{\infty}
\frac{1}{\alpha^2 + \kappa^2} d\kappa 
 =  \frac{\alpha \gamma}{8} \nn \\
g_{\gamma \gamma} & = & \frac{1}{8 \pi} 
\int_{k'} \left( \frac{\partial \theta_{k'}}{\partial \gamma} \right)^2 dk'
\nn  \\
& = & \frac{\alpha^2}{8 \pi \gamma} \int_{-\infty}^{\infty}
\frac{\kappa^2}{\left( \alpha^2 + \kappa^2 \right)^2} d \kappa 
 =  \frac{\alpha}{16 \gamma} ~.
\label{eq:g_phi_gamma}
\eeq
Finally, we use our earlier equation for the boundary
Euler integral to arrive at
\be
\chi_\mathrm{boundary} = -\frac{\partial \sqrt{g_{\phi \phi}} / 
\partial \gamma}{\sqrt{g_{\gamma \gamma}}} = 
-\frac{\frac{\sqrt \alpha}{4 \sqrt 2} \gamma^{-1/2}}
{\frac{\sqrt \alpha}{4} \gamma^{-1/2}} = -\frac{1}{\sqrt 2} ~.
\ee
Thus, for all models whose low-energy Hamiltonians are described by \eref{eq:Hk_general}, 
the bulk Euler integral between the anisotropic QCP and the
geodesic at $\gamma=\infty$ remains $\frac{1}{\sqrt 2}$.

\subsection{Robustness against angular distortions}
\label{sec:angular_distortions}

The previous section demonstrated robustness of the Euler integrals at phase 
transitions for the case where the metric is diagonal.  In
addition, while changes of coordinates can impact the critical scaling
properties (at least from a mathematical perspective), the conclusions that
we drew were with regards to geometric invariants, and thus manifestly
unaffected by such a coordinate change.
However, our physical intuition from the theory of continuous phase transitions
suggests that this robustness should be even more general, allowing arbitrary
perturbations to the model as long as they do not change the scaling properties
of the critical point (with regards to traditional observables).
Therefore, in this section we demonstrate that
perturbations which satisfy this constraint while
introducing off-diagonal components to the metric
nevertheless do not change the value of the Euler integrals.

In the $h-\phi$ plane, a simple method for introducing off-diagonal 
terms to the metric is to allow $\gamma$ to vary in the vicinity of the 
QCP.  Let $\gamma$ be some arbitrary
function $\gamma(h,\phi)$, with the restriction that $\gamma>0$
so that we remain in the same phase.  With this additional freedom, we
get a new metric $g'(h,\phi)$ such that
\beq
\nonumber
ds^2 & = & g_{hh} dh^2 + g_{\phi \phi} d\phi^2 + g_{\gamma \gamma} d\gamma^2
+ 2 g_{h \gamma} dh d\gamma \\
& = & g'_{hh} dh^2 + g'_{\phi \phi} d\phi^2 + 2 g'_{h \phi} dh d\phi ~.
\eeq
Noting that $d \gamma=(\partial_h \gamma) dh + (\partial_\phi \gamma) d \phi$, 
we find
\beq
g'_{hh} &=& g_{hh} + 2 g_{h \gamma} (\partial_h \gamma) + g_{\gamma \gamma} 
(\partial_h \gamma)^2 \nonumber \\
g'_{\phi\phi} &=& g_{\phi\phi} + g_{\gamma \gamma} (\partial_\phi \gamma)^2 \nonumber \\
g'_{h\phi} &=& g_{\gamma \gamma} (\partial_h \gamma) (\partial_\phi \gamma) + 
g_{h \gamma} (\partial_\phi \gamma) ~.
\eeq
Close to the critical point, only one term diverges: 
$g_{hh} \sim |h-h_c|^{-1} \implies g'_{hh} \sim |h-h_c|^{-1}$, while both $g'_{\phi \phi}$ and 
$g'_{h \phi}$ remain finite near the critical point. Thus, $g'$ is asymptotically
diagonal near the critical point, our earlier arguments still work, and
the boundary Euler integral remains zero.

Not surprisingly, the non-integer bulk Euler integral of the anisotropic phase transition
is more sensitive to details of the perturbation.  
For instance, the most naive option of giving the transverse
field a functional dependence ($h\to h(\gamma,\phi)$) changes the value of the bulk
Euler integral, which is not surprising given that $h$ is a relevant
perturbation at this phase transition. This can be traced back to the fact that modifying $h$ changes the 
position of the gapless momentum $k_0$ (see \eref{eq:Hkprime_anisotropic}), strongly affecting
the low-energy physics near the critical point.

In the absence of physical parameters to modify, we instead consider modifications
to the low-energy Hamiltonian.  In particular, consider a slightly more 
general Hamiltonian of the form:
\be
H_{k'} = \beta k' \sigma_{k'}^z + 
\beta \gamma
\left[ \alpha_x(\phi) \sigma^x_{k'} \cos \phi  + \alpha_y(\phi) \sigma^y_{k'} \sin \phi \right] ~,
\label{eq:conical_anisotropic}
\ee
where we demand that the functions $\alpha_{x,y}$ are periodic ($\alpha_{x,y}(0)=\alpha_{x,y}(2\pi)$)
and positive, such that the azimuthal Bloch angle still wraps the sphere once 
as we take $\phi$ from $0$ to $2\pi$.

To determine if the Euler integral $\chi_\mathrm{boundary}=-1/\sqrt{2}$ is protected, we
numerically solve for the boundary Euler integral for a variety of functions $\alpha_{x,y}$.  In doing so,
we require an additional constraint to ensure that this integral is well-defined: the 
metric must be positive definite, i.e., its determinant $g=EG-F^2$ must be non-zero.  
We have tested a number of functions satisfying these constraints, and found that
all of them have $\chi_\mathrm{boundary}=-1/\sqrt{2}$ as expected.  Given that the most complex 
functions we tested ($\alpha_x=1+\frac{\cos(\phi^2/\pi)}{4}$ and 
$\alpha_y=2+\sin \phi$) have no special symmetries, we postulate that the Euler
integral is identical for all functions satisfying the above constraints;
however, we are unable to analytically prove such a statement at this time.

\section{Classification of singularities}

Using scaling arguments, we have demonstrated the robustness of the geodesic curvature 
and the bulk Euler integral for situations where the boundary of the parameter 
manifold coincides with the phase boundary. One obvious difference
between the model in the transverse field ($h-\phi$) and the anisotropy ($\gamma-\phi$)
planes is integer vs. non-integer values of the Euler integrals. In this section 
we show how this difference comes from the nature of the singularities at the 
respective phase boundaries.  We identify two types of geometric singularity:
integrable singularities, as in the case of the $h-\phi$ plane, and conical
singularities, as in the case of the $\gamma-\phi$ plane.  Finally, in the 
$h-\gamma$ plane, we identify a third type of singularity, known as a curvature
singularity.  We discuss general conditions
under which these singularities should occur and, for the case of conical 
singularities, identify the relevant parameters in determining the boundary
Euler integral.

\subsection{Integrable singularities}
\label{sec:singularities_integrable}

A simple question which we must ask before classifying the geometric singularities of the 
XY chain are what, precisely, do we mean by singularities? A simple definition,
namely the divergence of one or more components of the metric tensor, 
is certainly a useful tool for diagnosing the presence of
phase transitions in practice \cite{Schwandt2009_1,Albuquerque2010_1}.
However, we claim that this singularity is
less fundamental from a geometric standpoint.  For instance, in the case of the 
TFI model, the transverse field component of the metric tensor diverges as
$g_{hh} \sim |h-1|^{-1}$ near $h=1$.  However, this divergence can be
removed by simply reparameterizing in terms of $h'=\sqrt{h-1} \sgn(h-1)$, for
which $g_{h' h'} \sim 1$\cite{Dey2012_1}.  Therefore, we need to look elsewhere for information
about the fundamental nature of the singularities in the quantum geometry.

Since the issue with the metric tensor was its coordinate dependence, natural
quantities to look at are the geometric invariants introduced in \sref{sec:geom_invariants},
which are coordinate-independent. Of these, the Gaussian curvature is 
the obvious choice \cite{Zanardi2007_1,Zanardi2007_2,Zanardi2007_3,Dey2012_1}.
We therefore classify singularities here and in the rest of the paper
based on the Gaussian curvature, $K$, and its invariant integral, $\chi_\mathrm{bulk}$.

For the case of the TFI model, the curvature does not diverge near the 
critical point.  This can be easily seen in the equivalent three-dimensional 
manifold (\fref{fig:ising_h_phi}), where the curvature goes from that of
a cylinder ($K=0$) to that of a sphere ($K=1/a^2$, where $a$ is the radius),
both of which are finite.  As we show more explicitly in \sref{sec:singularities_curvature},
one can derive this non-divergent result by
using the scaling forms of the metric tensor to get $K \sim \mathrm{const}$.  

However, critical scaling theory does not demand that the curvature
is a smooth function of the transverse field.  Indeed, we expect it to
be singular (like most other quantities) in the vicinity of a phase transition,
which manifests in the TFI chain as a jump of $K$ between the ferromagnet and the paramagnet.
However, the curvature is finite at all points, and is therefore completely
integrable when determining $\chi_\mathrm{bulk}$.  Therefore, we refer to
these jumps in the curvature as ``integrable'' singularities.  We note that visually these integrable
singularities correspond to points where the manifold changes shape locally,
but in such a way that the tangent plane evolves continuously, so that no
cusps or other points of curvature accumulation occur.

\subsection{Conical singularities}
\label{sec:singularities_conical}

The anisotropic phase transition at $\gamma=0$ is
an example of a conical singularity \cite{Fursaev1995_1, Kreyszig1959_1}, which can easily be seen
in \fref{fig:anising_g_phi}.  While the specific value of $1/\sqrt 2$
for the bulk Euler integral is likely specific to this particular class of models,
we claim that the existence of conical singularities is in fact a much more general
phenomenon.

More specifically, we expect conical singularities to occur in situations with 
two inequivalent directions orthogonal to a line (or a higher dimensional manifold) 
of critical points, as long as the orthogonal directions have the 
same scaling dimension. \footnote{We note that the parameters need not originally
behave identically if, by appropriate reparameterization, the couplings can
be made to have the same scaling dimensions.} Denote these directions 
$\lambda_1$ and $\lambda_2$, with the critical point at
$\lambda_1=\lambda_2=0$.  At $\gamma=0$ in the anisotropic XY
model, the parameter $\phi$ has no
effect on $H$, so in the FM phase this model satisfies the criteria for a 
conical singularity with $\lambda_1=\gamma \cos \phi$ and $\lambda_2=\gamma \sin \phi$. 

For simplicity we also assume that the metric has cylindrical symmetry, 
as in the case of the anisotropic transition in the XY model. 
In the previous section we verified numerically that this singularity, 
and in particular the boundary contribution to the Euler characteristic, is protected 
against breaking of the cylindrical symmetry.  We nevertheless use this
assumption to simplify our analysis. To ensure cylindrical symmetry,
the metric tensor should be diagonal in the $\lambda-\phi$ plane with the leading 
order asymptotic of the diagonal components of the metric tensor scaling as some power laws:
\be
g_{\lambda \lambda}=A \lambda^{-\alpha},~~g_{\phi \phi} = B \lambda^\beta ~,
\ee 
where $A$ and $B$ are arbitrary positive constants and we generally expect $\alpha \geq 0$, 
$\beta>0$.  However, if we define $\lambda_1=\lambda \cos \phi$ and 
$\lambda_2=\lambda \sin \phi$, then the demands of uniform 
scaling place an additional constraint on the values of the
scaling dimensions.  To see, this consider the components of the metric in 
``Cartesian coordinates'':
\beq
\nonumber
g_{11} &=& g_{\lambda \lambda} \left( \frac{\partial \lambda}{\partial \lambda_1} \right)^2
+ g_{\phi \phi} \left( \frac{\partial \phi}{\partial \lambda_1} \right)^2 \\
\nonumber
&=& g_{\lambda \lambda } \frac{\lambda_1^2}{\lambda^2} + 
g_{\phi \phi} \frac{\lambda_2^2}{\lambda^4} \\
\nonumber
&=& A \lambda^{-\alpha} \cos^2 \phi + B \lambda^{\beta-2} \sin^2 \phi \\
\nonumber
g_{22} &=& A \lambda^{-\alpha} \sin^2 \phi + B \lambda^{\beta-2} \cos^2 \phi \\
g_{12} &=& \left( A \lambda^{-\alpha} + B \lambda^{\beta - 2} \right) \cos \phi \sin \phi~.
\label{eq:g11_g22}
\eeq
We clearly see that the scaling dimensions of $\lambda_1$ and $\lambda_2$ are 
the same at all angles $\phi$ if and only if the 
exponents satisfy the relation:
\be
\beta = 2 - \alpha.
\label{eq:beta_alpha}
\ee 
Note that the condition $\beta>0$ now means that $0 \leq \alpha < 2$. The constants 
$A$ and $B$ are non-universal, but we expect that their ratio $B/A$ -- which defines the 
anisotropy of the metric tensor -- will be a universal number for a given
class of models. For the anisotropic transition 
this ratio is $B/A=2$ and the exponents are $\alpha=\beta=1$ (see \eref{eq:g_phi_gamma}).
Interestingly the point $h=\infty$ in the $h-\phi$ plane (corresponding the 
spherical cap -- see \fref{fig:ising_h_phi}) also has the form of a
conical singularity if we use $\lambda_1=\frac{1}{h} \cos(\phi)$ and
$\lambda_2=\frac{1}{h}\sin(\phi)$ with $\alpha=0$, $\beta=2$, and $B/A=1$. 
These exponents describe a non-singular point in 
parameter space with cylindrical symmetry.

\begin{figure}
\includegraphics[width=.8\linewidth]{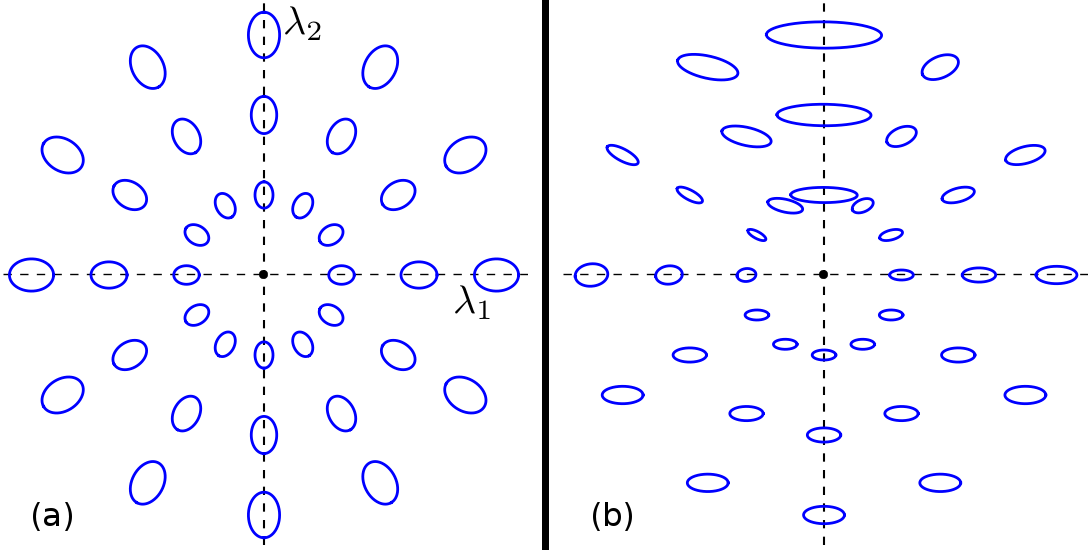}
\caption{Graphical representation of the metric in the $\lambda_1$-$\lambda_2$
plane, where a conical singularity (critical point) is at $\lambda_1=\lambda_2=0$.  Panel
(a) shows the simple case of the XY model in a transverse field, for which 
$\alpha=\beta=1$, $B/A=2$, and the metric is cylindrically symmetric about the 
origin. Panel (b) shows the more general case with Hamiltonian given by 
\eref{eq:conical_anisotropic}, in which the cylindrical symmetry has been broken
by using the functions $\alpha_x(\phi)=1+\frac{1}{2} \cos(2 \phi) - \frac{1}{3} \sin(2 \phi)$
and $\alpha_y(\phi)=2+\sin(\phi)$.  While the cylindrical symmetry is gone, we numerically
find that this model has the same conical angle as when $\alpha_x=\alpha_y=\mathrm{const}$,
yielding the same bulk Euler integral.
The metric is plotted in both panels by showing the ``shape of the circle'' near each point, i.e.,
the blue ellipses show contours of constant radius in the $\lambda_1$-$\lambda_2$ plane.
Since the metric diverges near the critical point, the size of the ellipses gets smaller.}
\label{fig:conical}
\end{figure}

Given a conical singularity, we can now easily find the Euler integral.  Using the
same formulas as earlier for the case of cylindrical symmetry,
\beq
\chi_\mathrm{boundary} &=& -\frac{\partial \sqrt{g_{\phi \phi}}/\partial \lambda}
{\sqrt{g_{\lambda \lambda}}} \\
&=&\left(\frac{\alpha}{2} - 1\right) \sqrt{\frac{B}{A}} ~.
\label{eq:chi_bound}
\eeq
Using this formula for  the anisotropic phase transition of the 
XY model yields $\chi_\mathrm{boundary}=-1/\sqrt 2$,
as found earlier.  For a demonstration of the contours of the metric in this model, see
\fref{fig:conical}.

Let us point that if we have a non-singular point like $h\to\infty$ at fixed $\gamma$, for which $\alpha=0$ 
and $\beta=2$, we have an additional requirement that $g_{11} = g_{22}$ and $g_{12}= 0$, implying
that $A=B$. This follows from the fact that the metric must remain regular at $\lambda=0$. 
Then from \eref{eq:chi_bound}  we find that the boundary contribution of the isotropic 
point will be $\chi_\mathrm{boundary}=-1$.  In a three-dimensional embedding, this 
indeed looks like a hemisphere, which is non-singular.  Therefore, for such smooth 
``singularities,'' the manifold is also guaranteed to be locally equivalent to the hemisphere.

Finally, for the case of a multi-critical point, one can try to apply similar logic. 
However, due to the asymmetry in the scaling dimensions, along some
direction the metric will be infinitely anisotropic near the critical
point. This infinite anisotropy is not generally removable by rescaling the couplings. 
Therefore, the conical singularity breaks down and the curvature can become non-integrably singular.

\subsection{Curvature singularities}
\label{sec:singularities_curvature}

\begin{figure}
\includegraphics[width=\linewidth]{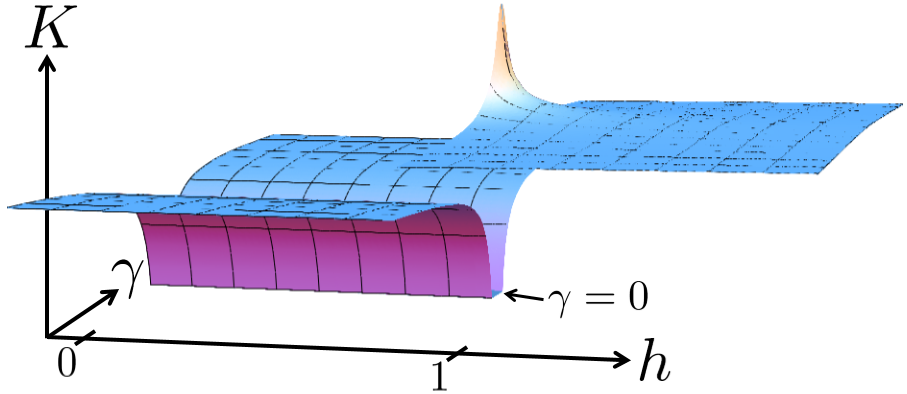}
\caption{
Three dimensional plot of $K(h,\gamma)$, similar to that found in 
Fig. 1 of \rcite{Zanardi2007_1}. The graph shows 
curvature singularities along the two phase transitions $h=1$ (only positive $h$ are shown) 
and $h<1$, $\gamma=0$. Unlike the Ising phase transition, the curvature 
singularities near the anisotropic phase transition are non-integrable, 
leading to divergent Euler integral within each phase.}
\label{fig:curvature_h_gamma}
\end{figure}

If we now consider the third two-dimensional cut of the XY model, 
namely the $h-\gamma$ plane (which has been
solved for previously in \rcite{Zanardi2007_2}),
we find that the curvature displays a number of additional singularities.
The structure of these singularities can be seen in~\fref{fig:curvature_h_gamma}. 
It is clear from the plot that as expected there are singularities near the two phase transitions: 
integrable singularities near the Ising transition ($|h|=1$) and non-integrable singularities near 
the anisotropic transition ($|h|<1$ and $\gamma=0$). These singularities meet near the 
multi-critical point ($|h|=1$ and $\gamma=0$) resulting in a very singular and non-monotonic 
behavior of the curvature. 

While, unlike the $h-\phi$ and $\gamma-\phi$ planes,
there are no obvious finite protected Euler integrals in the
$h-\gamma$ plane, the exponent of the curvature divergence can be 
found from scaling arguments.  For instance, near the Ising phase transition ($|h|=1$,
$\gamma > 0$), the simple arguments of Venuti and Zanardi \cite{Venuti2007_1} 
indicate that the metric components diverge as $g_{hh} \sim |h-h_c|^{-1}$,
$g_{\gamma \gamma} \sim 1$, and $g_{h \gamma} \sim 1$. The non-divergence of
$g_{\gamma \gamma}$ is due to the fact that $\gamma$ is a marginal parameter,
such that the scaling dimension of the singular part of $g_{\gamma \gamma}$ is $+1$.
However, there is also a non-zero non-singular part, which is the leading order
term near the critical point.  Similarly, the scaling dimension of 
$g_{h \gamma}$ is zero, such that there is a jump singularity near the 
critical point.  Plugging these divergences into \eref{eq:invariants} and
assuming a smooth dependence on $\gamma$ as long as we are far from the
multi-critical point, we find that $K \sim 1$, which matches with the jump singularity in
$K$ found at the Ising phase transition.  By contrast, near the anisotropic
phase transition ($\gamma=0$, $|h| < 1$), both the transverse field $h$ and the 
anisotropy $\gamma$ are relevant parameters, with scaling dimension $-1$.
Therefore, the metric components scale as $g_{hh} \sim g_{h\gamma} \sim 
g_{\gamma \gamma} \sim |\gamma|^{-1}$.  By the same logic as before, this
leads to a divergent curvature with $K \sim |\gamma|^{-1}$.

Finally, we point out some subtleties regarding the curvature far from 
critical points.  First, while the curvature does not diverge near $h=\infty$ or $\gamma=\infty$, it is
not immediately apparent whether the bulk Euler integral diverges in this limit.  Therefore,
in \aref{sec:reparam}, we show that a simple reparameterization allows one to 
make both $K$, $\sqrt g$, and the limits of integration simultaneously finite, except
near quantum critical points at finite $h$ and $\gamma$.  Therefore, it becomes
clear that the divergence in the Euler integrals comes strictly from the curvature
singularities near the quantum phase transitions.

Second, we note that the metric component $g_{hh}$ vanishes
at $\gamma=0$ for all $h>1$, since the ground state is the fully-polarized
spin-up state along this entire line.  The general intuition
from Riemann geometry is that if the determinant of the metric vanishes, then
the curvature diverges, since the determinant appears in the denominator of the 
curvature formula (\eref{eq:invariants}). However, as we show in the next
section,  the curvature does not in fact diverge for fundamental physical reasons.

\subsection{Metric singularity near lines of symmetry}
\label{sec:K_symm_line}

Consider the line $|h|>1$, $\gamma=0$ in the XY model.  The
ground state along this entire line is fully polarized along the direction
of the transverse field, which is clearly the ground state at
$h=\infty$.  Then, since along this special XY-symmetric
line the Hamiltonian commutes with $S^z_\mathrm{total}$, the fully
polarized eigenstate must remain the ground state until a gap closes.
Note that this argument continues to hold even in the presence of
integrability-breaking perturbations, as long as the $z$-magnetization
remains a good quantum number.  Therefore, such a line of unchanging
ground states and thus vanishing metric determinant is a robust feature
of this class of models.  

More generally, one can create such fully polarized ground states
by considering a family of the Hamiltonians
\be
\mathcal H(\lambda,\delta)=\mathcal H_0(\lambda,\delta)-\lambda  \mathcal M~,
\ee 
where $\delta$ is a symmetry breaking field such that $\mathcal H_0$
and $\mathcal M$ commute at $\delta=0$ for any value of $\lambda$. Physically
$\mathcal M$ represents the generalized force ($z$-magnetization in the above example)
which is conserved at the symmetry line. In general $\mathcal M$ can also 
depend on $\lambda$ and $\delta$ as long as $\mathcal H_0$ and $\mathcal M$ 
commute at $\delta=0$. Clearly in the limit $\lambda\to\infty$ the ground state 
of the Hamiltonian is the fully polarized state (the state with largest eigenvalue) 
of the generalized force $\mathcal  M$, which is generally non-degenerate. 
By the argument above, along the symmetry line the ground state of $\mathcal H$ will 
be independent of $\lambda$ until the gap in the Hamiltonian closes, e.g., until the 
system undergoes a quantum phase transition. Thus the metric near this symmetry 
line will be singular with a vanishing determinant. \footnote{We point out that having symmetry 
is not sufficient to get a vanishing metric. If the ground state of the Hamiltonian at 
$\delta=0$ corresponds to a degenerate eigenstate of $\mathcal M$, then it is 
not protected against small changes in $\lambda$ and the metric is generally non-singular.}

We now investigate why, despite this ``singular'' metric, the curvature 
nevertheless remains analytic in the vicinity of such a line of symmetry. 
Assuming we are far from any critical points (i.e., with a gapped spectrum), 
the components of the metric tensor near the fully polarized state should be analytic and can be written as
\be
g=\left(
\begin{array}{cc}
\delta^2  f_{\lambda\lambda} & \delta f_{\delta\lambda}\\
\delta f_{\delta\lambda} & f_{\delta\delta} 
\end{array} \right) ~.
\label{eq:tens_symm_point}
\ee
All components $f_{ij}$ are smooth functions of $\lambda$ and $\delta$. 
For small $\delta$ they can be approximated as being independent 
of $\delta$, $f_{ij} = f_{ij}(h)$, since the leading asymptotic of the curvature 
in the limit $\delta\to 0$ 
will be determined by the explicit dependence given by \eref{eq:tens_symm_point}.

Using the explicit expression for the curvature (\eref{eq:invariants}) and 
counting powers of $\delta$, we see that the only possible divergent term in the curvature 
as $\delta \to 0$ is given by:
\be
K\approx -{1\over \delta^2 \sqrt{f}} {\partial\over \partial\lambda}\left({\sqrt{f}\over f_{\lambda\lambda} }\Gamma^\delta_{\lambda\delta}\right),
\ee
where $f=\mathrm{det}(f_{ij})$ and 
\be
\Gamma^{\delta}_{\lambda\delta}={1\over 2g}\left[g_{\lambda\lambda} {\partial g_{\delta\delta}\over \partial \lambda}-g_{\delta\lambda} {\partial g_{\lambda\lambda}\over \partial\delta}\right]={f_{\lambda\lambda} \over 2f}\left[ {\partial f_{\delta\delta}\over \partial\lambda}-2 f_{\delta\lambda}\right].
\ee
Therefore
\be
K \approx -{1\over 2\delta^2\sqrt{f}} {\partial\over \partial \lambda}\left[{\partial_\lambda f_{\delta\delta}-2 f_{\delta\lambda}\over \sqrt{f}} \right].
\label{sing_curvature}
\ee

In order to see that the curvature is not divergent near this line of
symmetry, we need to show that the following difference vanishes
\be
\partial_\lambda f_{\delta\delta}-2 f_{\delta\lambda}=\partial_\lambda g_{\delta \delta}-2 \partial_\delta g_{\lambda \delta}=0.
\label{eq:metric_symm_point}
\ee
This is indeed the case for such general lines of symmetry, as we prove in detail in
\aref{sec:appendix_proof}. Therefore, the singular term in the curvature vanishes, 
and thus the curvature does not diverge near the symmetric line.  We 
emphasize again that our conclusions regarding the curvature are invariant
under reparametrization of the couplings.

A more geometric view of the absence of a curvature singularity for metrics 
satisfying \eref{eq:metric_symm_point} can be seen by mapping this metric 
to a Euclidean plane such that all distances are preserved (i.e., an 
isometric mapping). We start by switching back to the original couplings
of the XY-model, $\lambda\leftrightarrow h$ and $\delta\leftrightarrow \gamma$. Then
\beq
\nn ds^2 &=& \gamma^2 f_{hh} d h^2 + \gamma \partial_h f_{\gamma \gamma} 
d h d \gamma + f_{\gamma\gamma} d \gamma^2 
\\ &=& r^2 d\varphi^2 + dr^2,
\eeq
where to the leading order in $\gamma$
\be
f_{hh}=\frac{h}{16 (h^2-1)^{5/2}},\;
f_{\gamma \gamma}=\frac{h-\sqrt{h^2-1}}{8\sqrt{h^2-1}}.
\label{eq:f_hh_gg}
\ee

We use the natural ansatz in polar coordinates:
\be
r=r(h,\gamma)~,~\varphi=\varphi(h) ~.
\ee
This gives the metric
\beq
\nn ds^2&=&r^2 (\partial_h \varphi)^2 dh^2 + (\partial_h r)^2 dh^2+
\\&& 2 (\partial_h r)(\partial_\gamma r) dh d\gamma + 
(\partial_\gamma r)^2 d\gamma^2 ~.
\eeq
Matching the $d\gamma^2$ terms, we get $\partial_\gamma r = 
\sqrt{f_{\gamma\gamma}}$, implying that $r=\gamma \sqrt{f_{\gamma \gamma}}$.  
This also works to match the $dh d\gamma$ terms:
\be 
2 (\partial_h r)(\partial_\gamma r) = 2 \gamma \sqrt{f_{\gamma \gamma}} \partial_h \sqrt{f_{\gamma \gamma}}
 = \gamma \partial_h f_{\gamma \gamma} ,
\ee
demonstrating the importance of \eref{eq:metric_symm_point} to obtain a flat metric.
Finally, matching the $dh^2$ terms, we get
\begin{align}
\nn \gamma^2 f_{h h} &= r^2 (\partial_h \varphi)^2 + (\partial_h r)^2 
\\ \nn  &= \gamma^2 (\sqrt{f_{\gamma \gamma}})^2 (\partial_h \varphi)^2 + \gamma^2 (\partial_h \sqrt{f_{\gamma \gamma}})^2 
\\ \varphi  &= \int_h^\infty \frac{dh'}{\sqrt{f_{\gamma \gamma}(h')}} 
\sqrt{f_{h h} (h')
 - (\partial_{h'} \sqrt{f_{\gamma \gamma}(h')})^2},
\end{align}
where the limits of integration have been chosen to give $\varphi>0$ for 
all $h$ and $\varphi(h\to\infty) \to 0$. 
The map is well-defined as long as the term in the square root
is positive, i.e., as long as $f_{hh}-(\partial_h \sqrt{f_{\gamma\gamma}})^2 > 0$.
It's easy to check that this difference is indeed positive:
\be
f_{hh} - (\partial_h \sqrt{f_{\gamma\gamma}})^2 = 
\frac{h-\sqrt{h^2-1}}{32(h^2-1)^{5/2}} ~.
\ee
Hence, this embedding works, and shows that the surface for this simplified
metric is equivalent to a plane.  As such, the curvature is easily seen to be zero. 

\begin{figure}
\includegraphics[width=\linewidth]{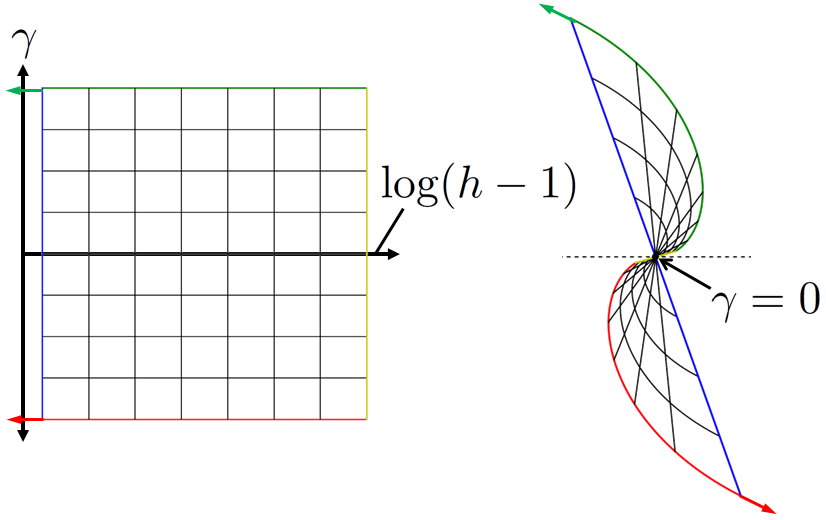}
\caption{Mapping of the curved space in the $h-\gamma$ plane to a 
flat plane in polar coordinates.  A grid covering a  subset of the $h-\gamma$
plane (left) is shown mapped to a subset of the flat plane (right) by a 
coordinate change to $r$ and $\varphi$ as described in the
text.  The mapping is specifically shown for the leading
order asymptotics  of the metric of the  XY model given by Eqs.~\ref{eq:f_hh_gg}, 
for $h$ ranging from $1.0009$ to $2.0$ and $\gamma$ from $-0.1$ to $0.1$.  
The red and green arrows indicate that both $\varphi$ and $r$ diverge
in the limit $h\to 1^+$. Since $g_{hh}=0$ for $\gamma=0$, 
the entire line $\gamma=0$ maps 
to a single point, which is a  singularity of the model.}
\label{fig:plane_mapping}
\end{figure}

While this mapping shows that the manifold in the $h-\gamma$ plane is much
less singular than expected, there is still in some sense a singularity at the line of 
symmetry.  This can be see in \fref{fig:plane_mapping}, where the entire line $\gamma=0$ 
maps to a single point in the $r-\varphi$ representation.  The singularity does not 
show up directly as a divergence in the scalar curvature.  But 
consider the line $\gamma=0$, and note that any embedding
into a higher-dimensional flat space must identify all the points
on this line, since $g_{hh}=0$.  At the same
time, the curvature $K$ is not independent of $h$ along this line (see 
\fref{fig:curvature_h_gamma}).  Therefore, the curvature
cannot be a smooth function for such an embedding, since the curvature upon 
approaching the point $\gamma=0$ will depend on the direction in which it is approached.
By a similar logic, the curvature is singular in the $h-\phi$ plane at $\gamma=0$, diverging
as $1/\gamma^2$.  This is visible in the simple three-dimensional embedding discussed in 
\sref{sec:visualizing}; as $\gamma \to 0$, the radius of the hemisphere decreases to zero,
and thus the curvature diverges.   Finally, as we
show in detail in \aref{sec:curv_3d}, these two divergences conspire to cause
the scalar curvature (Ricci scalar) for the full three-dimensional
manifold to diverge along this line of symmetry\footnote{We thank
one of the referees for bring this to our attention.}.  This divergence is remarkable since the symmetry line does not formally correspond to any phase transition.  However, the physical 
interpretation of these divergences remains unclear to us at this time.

\section{Measuring the metric}
\label{sec:measuring_metric}

Formally the metric tensor can be expressed as a standard response function~\cite{Venuti2007_1} and 
thus is in principle measurable. However, unlike the Berry curvature, which naturally appears in the 
off-diagonal Kubo-type response~\cite{DeGrandiUnpub2013_1}, the metric tensor appears either as a 
response in imaginary time dynamics~\cite{DeGrandi2011_1,DeGrandiUnpub2013_1} or as a 
response in dissipative systems using specific -- and usually not physically justified --
requirements for the dissipation~\cite{Avron2011}. However, as shown in \rcite{NeupertArxiv2013_1} 
for the specific situation of non-interacting particles, the geometric tensor 
characterizing the Bloch bands
can be measured through the spectral function of the current noise. Here we extend this idea 
to arbitrary systems and couplings.

The geometric tensor can be represented as 
(cf. Eq. (B8) in \rcite{DeGrandi2011_1})
\be
\chi_{\mu\nu}=\int {d\omega\over 2\pi} {S_{\mu\nu}(\omega)\over \omega^2},
\label{eq:metric_corr_fn}
\ee
where
\be
S_{\mu\nu}(\omega)=\int_{-\infty}^\infty dt \mathrm e^{-i\omega t}\bra \partial_\mu H(t) \partial_\nu H(0)\ket_c
\ee
is the Fourier transform of the connected ground state non-equal time correlation function of the generalized forces, 
a.k.a. the noise spectral density, and $\partial_\mu H(t) = e^{i H t} \partial_\mu H(0) e^{-i H t}$ is the
generalized force in the Heisenberg picture. The metric tensor is the symmetric (real) 
part of the  geometric tensor and thus can be expressed through the symmetrized spectral density. 
Following the insight of Neupert et al.~\cite{NeupertArxiv2013_1}  we interpret $S_{\mu\nu}(\omega)$ 
as the Fourier transform  of the non-equal time noise correlation function of two generalized forces,  
which is relevant  experimentally~\cite{Blanter2000_1}.  For example, in mesoscopic systems,
the current noise spectrum $S_{\mu \nu}(\omega)$ can be measured in shot noise experiments, where 
current $J_\mu$ corresponds to the generalized force $\partial_{A_\mu} H$. 
\eref{eq:metric_corr_fn} suggests a simple and general way of measuring the metric tensor 
in interacting many-body systems by analyzing equilibrium noise. We believe that, for sufficiently large
systems, such symmetrized noise correlations should be measurable 
with negligible effects of measurement back-action.

While the method of measuring the metric tensor through noise is conceptually simple, it cannot be 
easily implemented in systems such as cold atoms, where measurements are often destructive. 
Below, we discuss two real-time protocols which offer the possibility of observing the metric tensor 
via destructive (single-time) measurements.  We then mention an additional protocol involving instantaneous 
quenches of the external parameters.  We note that both ramps \cite{Simon2011_1,Kim2011_1} and quenches
\cite{Greiner2002_2,Sadler2006_1,Tuchman2006_1} are routinely achieved in isolated cold atom systems.

Consider performing real-time ramps of some parameter $\lambda_\mu$ in a gapped system,
starting from the ground state at the starting point $\lambda_i$.  
It has been shown elsewhere \cite{DeGrandiUnpub2013_1} that,
for a square root ramp with $\lambda_\mu(t)-\lambda_\mu(t_f) = \left[ v (t_f-t) \right] ^{1/2}$, 
the leading order correction to the energy in the limit $v \to 0$ is
given by
\be
\bra H \ket = E_0 + v g_{\mu \mu} + O(v^2) ~,
\ee
where $H$ is the Hamiltonian, $E_0$ is its ground state energy, and
$g_{\mu \mu}$ is diagonal component of metric along the ramping direction.


However, the square root ramp is singular near $t_f$, and therefore may
be difficult to implement.  We now show that the metric can also be measured via
a more easily implemented linear ramp, at the cost of requiring a harder
measurement: the quantum energy fluctuations. 

Consider a linear ramp $\lambda_{\mu}(t) - \lambda_\mu (t_f) = v (t-t_f)$. 
From \rcite{DeGrandi2010_2}, we know that the wave function 
at $t_f$ will be given in its instantaneous eigenbasis by
\be
|\psi\ket = (1+\beta v^2) |0\ket - i v \sum_{n \neq 0} \alpha_n |n\ket + O(v^2)~,
\ee
where $\alpha_n = \frac{\bra n | \partial_\mu H | 0 \ket}{(E_n - E_0)^2}$ and
$\beta=-(1/2)\sum_{n \neq 0} |\alpha_n|^2$, 
which serves to keep the wave function normalized
up to order $v^2$. The energy fluctuations are given by
$\Delta E^2 = \bra \psi | H^2 | \psi \ket - \bra \psi | H | \psi \ket ^2$.
Without loss of generality, we may offset the Hamiltonian such
that the ground state energy is $E_0=0$.  Then, up to order $v^2$,
\beq
\nn
\Delta E^2 &=&
v^2 \sum_{n \neq 0} |\alpha_n|^2 \bra n | H^2 | n \ket \\
\nn
&=& v^2 \sum_{n \neq 0} |\alpha_n|^2 E_n^2 \\
&=& v^2 \sum_{n \neq 0} \frac{\bra n | \partial_\mu H | 0 \ket 
\bra 0 | \partial_\mu H | n \ket}{E_n^2} = v^2 g_{\mu \mu}~.
\eeq
Therefore, by measuring the energy fluctuations for different ramp rates
and extracting the leading order (quadratic) term, we can extract
diagonal terms of the metric along a given direction. Let us point that if we start the ramp in the 
ground state then the energy fluctuations are equal to the work fluctuations, so 
the metric tensor can be extracted by measuring work fluctuations as a function of the ramp rate.

A third possibility for measuring the metric tensor is by measuring
the probability of doing non-zero work for small quenches in
parameter space.  This is in some sense true by definition: 
if $|\psi_0(\lambda)\ket$ is
the ground state manifold, then the probability of doing zero work (i.e.
ending up in the ground state) after a quench from $\lambda_\mu$ to 
$\lambda_\mu+d \lambda_\mu$ is just
\beq
P(W=0)&=&\big|\bra \psi_0(\lambda_\mu) | \psi_0(\lambda_\mu+d\lambda_\mu) 
\ket\big|^2 \nonumber \\
&=&1-g_{\mu \mu} d \lambda_\mu^2 ~; \nonumber\\
P(W\neq0)&=&g_{\mu \mu} d \lambda_\mu^2 ~.
\eeq
As noted elsewhere, this quantity is equivalent to the time-averaged return 
amplitude G(t)~\cite{Silva2008_1}: 
\begin{eqnarray*}
P(W=0) & = & \lim_{T\to\infty}{1\over T}\int_0^T G(t) dt ~,\\
G(t) & = & \bra \psi_0(\lambda_\mu)|e^{i H(\lambda_\mu) t} 
e^{-i H(\lambda_\mu+d \lambda_\mu) t} |\psi_0(\lambda_\mu)\ket ~,
\end{eqnarray*}
which is related to the well-known Loschmidt echo $L(t)$ by
\be
L(t) = |G(t)|^2 ~.
\ee
The Loschmidt echo is also the probability of returning to the ground state 
(doing zero work) after a double quench of duration $t$ from $\lambda_\mu$ to 
$\lambda_\mu+d \lambda_\mu$ and back~\cite{HeylUnpub2012_1}.
While energy distributions and the related Loschmidt echo are in principle
measurable by a variety of methods, we note that there has been important 
recent progress in proposing measurements of these quantities using
few-level systems as a probe \cite{Heyl2012_1,MazzolaArxiv2013_1,DornerArxiv2013_1}.

Finally, we point out that one can reconstruct the full metric 
tensor solely from measurements of its diagonal components.  Consider a two-parameter manifold 
$(\lambda_x, \lambda_y)$.  First, measure the diagonal components $g_{xx}$ and 
$g_{yy}$ using one of the procedures described above.  Second, measure a specific 
off-diagonal element by varying $\lambda_x$ and $\lambda_y$ simultaneously.  For example, if we define the variable $\lambda_w=(\lambda_x+\lambda_y)/2$ and ramp or quench along the line $\lambda_x = \lambda_y$, we can obtain
$g_{ww}$.  Finally, noting that for this protocol $d\lambda_x=d\lambda_y=d\lambda_w$, we see that
\beq
ds^2=g_{ww} d\lambda_w^2 &=& g_{xx} d\lambda_x^2 + 2 g_{xy} d\lambda_x d\lambda_y + g_{yy} d\lambda_y^2 \nn \\
= g_{ww} d\lambda_x^2 & = & \left( g_{xx} + 2 g_{xy} + g_{yy} \right) d\lambda_x^2 \nn \\
\implies g_{xy} & = & \frac{g_{ww} - g_{xx} - g_{yy}}{2} ~.
\label{eq:gxy_meas}
\eeq
This procedure can be easily generalized to an $N$-parameter manifold by performing
pairwise measurements using a similar tricks as above.

\section{Conclusions}
\label{sec:conclusions}

In conclusion, using the quantum XY model as an example, we have analyzed
the Riemann manifold of a simple ground state phase diagram.  We identified
a new geometric characteristic -- the bulk Euler integral -- which characterizes
the phases of matter.  Based on the value of this Euler integral, either integer,
non-integer, or undefined, we have classified three types of singularities in the
Gaussian curvature: integrable, conical, and curvature singularities.  We showed
that integrable singularities occur for phase transitions where one parameter is
marginal or irrelevant while the other is relevant.  Similarly, conical singularities emerge when
the phase transition occurs at a single critical point with two ``orthogonal''
relevant directions that have the same scaling dimensions. And finally near the multi-critical point 
with two inequivalent relevant directions we found curvature singularities which, similar to black holes,
are non-removable non-integrable singularities in the quantum metric space.
Finally, by introducing additional techniques for measuring the metric experimentally, 
we point out that this geometric information should
be experimentally accessible.

\acknowledgments The authors would like to acknowledge useful and stimulating discussions with 
G. Bunin, C. Chamon, L. D'Alessio, and P. Mehta. We would also like to acknowledge one 
of our referees for bringing the divergence in the three-dimensional
scalar curvature to our attention. This work was partially supported by BSF 2010318, 
NSF DMR-0907039, NSF PHY11-25915, AFOSR FA9550-10-1-0110, the Swiss NSF, as well as the Simons and the Sloan 
Foundations. The authors thank the hospitality of the Kavli Institute for Theoretical Physics 
at UCSB and the support under NSF PHY11-25915.

\appendix

\section{Proof of \eref{eq:metric_symm_point}}
\label{sec:appendix_proof}

We now prove that \eref{eq:metric_symm_point} indeed holds for the metric 
given by \eref{eq:tens_symm_point} near the symmetric line. We will rely on 
the fact that, at this symmetric line, the ground state does not depend on $\lambda$, 
so $\partial_\lambda |0\ket=0$. Thus,
\begin{widetext}
\be
\partial_\lambda g_{\delta \delta}=\partial_\lambda \sum_{n\neq 0} 
\bra 0|\overleftarrow{\partial_\delta}|n\ket\bra n|\partial_\delta|0\ket=
\sum_{n\neq 0}\Bigg[ \left(  \bra 0|\overleftarrow{\partial^2_{\lambda\delta}}|n\ket+
 \bra 0|\overleftarrow{\partial_\delta}\partial_\lambda|n\ket\right)\bra n|\partial_\delta|0\ket+
\bra 0|\overleftarrow{\partial_\delta}|n\ket\left( \bra n|\overleftarrow{\partial_\lambda}
\partial_\delta|0\ket+\bra n|\partial^2_{\lambda \delta}|0\ket\right) \Bigg].
\ee

\beq
\nn
2\partial_\delta  g_{\delta \lambda}&=&\partial_\delta \sum_{n\neq 0} \Bigg[ \bra 0|\overleftarrow{
\partial_\delta}|n\ket\bra n|\partial_\lambda|0\ket+\bra 0|\overleftarrow{
\partial_\lambda}|n\ket\bra n|\partial_\delta|0\ket \Bigg] \\
&=&\sum_{n\neq 0}\Bigg[ \left(
  \bra 0|\overleftarrow{\partial^2_{\lambda\delta}}|n\ket+\bra 0|\overleftarrow{
\partial_\lambda}\partial_\delta|n\ket\right)\bra n|\partial_\delta|0\ket+
\bra 0|\overleftarrow{\partial_\delta}|n\ket\left( \bra n|\overleftarrow{
\partial_\delta}\partial_\lambda|0\ket+\bra n|\partial^2_{\lambda \delta}|0\ket\right) \Bigg]
\eeq
Therefore, we now find that \eref{eq:metric_symm_point} holds:
\beq
\nn
\partial_\lambda g_{\delta \delta}-2 \partial_\delta g_{\delta \lambda}&=&\sum_{n \neq 0} \Big[
 \bra 0|\overleftarrow{\partial_\delta}\partial_\lambda|n\ket\bra n|\partial_\delta|0\ket+
\bra 0|\overleftarrow{\partial_\delta}|n\ket\bra n|\overleftarrow{\partial_\lambda}
\partial_\delta|0\ket \Big] \\
&=&\sum_{n \neq 0,m} \Big[ \bra 0|\overleftarrow{\partial_\delta}
|m\ket\bra m|\partial_\lambda|n\ket\bra n|\partial_\delta|0\ket+
\bra 0|\overleftarrow{\partial_\delta}|n\ket\bra n|\overleftarrow{
\partial_\lambda}|m\ket \bra m|\partial_\delta|0\ket \Big] =0 ~.
\eeq
\end{widetext}
The last equality follows by observing that if we interchange indices 
$n$ and $m$ in the second term in the last sum, we get a term that exactly cancels 
the first one. This follows from 
\be
\bra m|\overleftarrow{\partial_\lambda}|n\ket=-\bra m|\partial_\lambda|n\ket.
\ee
In addition, the $m=0$ term vanishes because
\be
\bra 0 | \partial_\lambda | n \ket = - \bra 0 | \overleftarrow{\partial_\lambda} | n \ket = 0 ~.
\ee

\section{Choice of parameters}
\label{sec:reparam}

The goal of this section is to show by example that, if the bulk Euler integral
$2\pi \chi_\mathrm{bulk} = \int K dS$ diverges, then that implies that the curvature must
diverge at some point.  This is not a priori obvious, because the invariant area
$dS=\sqrt g d \lambda_1 d\lambda_2$ can also diverge, either because the metric
diverges or because the metric is finite but the parameters $\lambda_i$ have
infinite range.  We show that, for the $h-\gamma$ plane of the XY model, such
divergences can be removed by a suitable choice of coordinates.  

One natural thing to attempt to do is the 
go to ``unitless'' coordinate systems, $dh \to \sqrt{g_{hh}} dh$ and $d\gamma \to \sqrt{g_{\gamma 
\gamma}} d\gamma$, in which a one-parameter metric $g_{\lambda \lambda}$ 
would become flat (i.e., become $\lambda$-independent).  While this does not quite work the same for
two parameters, since $g_{hh}$ depends on $\gamma$, 
we nevertheless use a variant of it below to get a more well-behaved metric.  As we will see, the
new parameters have a finite range, and the metric is much more well-behaved.

Consider first the case of the transverse field $h$.  We wish to define a new parameter $\xi$ such that
$d\xi = \sqrt{g_{hh}} dh$.  The leading order $h$ dependence of $g_{hh}$ is $g_{hh} \sim \frac{1}{(1-h^2)}$
in the ferromagnet and $g_{hh} \sim \frac{1}{h^2 (h^2-1)}$ in the paramagnet.  Integrating these expressions
gives the natural choice
\be
\sin \xi \equiv \begin{cases} h & \mbox{if } |h| < 1 \\ 1/h & \mbox{if } |h| > 1~, \end{cases}
\ee
with the quadrant of $\xi$ chosen such that $|\xi| \in [0,\pi/2]$ if $|h| \leq 1$ and 
$|\xi| \in [\pi/2,\pi]$ if $|h| \geq 1$.  We similarly reparameterize the $\gamma$ direction in terms of
$\eta$ such that $d \eta = \frac{1}{\sqrt{\gamma} (1+\gamma)} d \gamma$, giving
\be
\eta \equiv 2 \tan^{-1} (\sqrt {|\gamma|}) \sgn \gamma ~. 
\ee
The range of our new, auxiliary variables is $\xi\in (-\pi,\pi)$, $\eta \in [0,\pi)$.

\begin{figure}
\includegraphics[width=\linewidth]{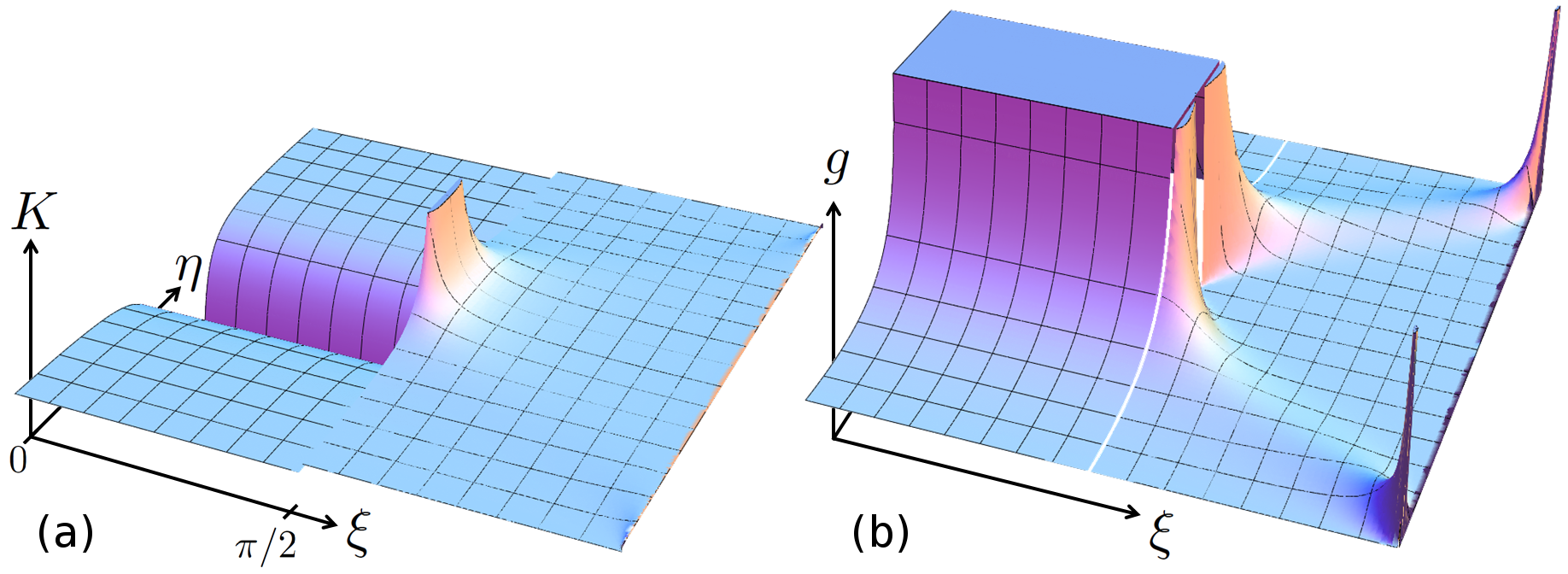}
\caption{
Geometric invariants of the XY model in the $\xi$-$\eta$ plane.  The
three dimensional plots show Gaussian curvature $K$ (a) and 
metric determinant $g$ (b) for the entire (finite) range of 
parameters $\xi$ and $\eta$.  As discussed in the text, the 
invariants clearly only diverge near the critical line at
$\gamma=0$, except for the determinant $g$, which has an
integrable divergence near $\xi=\eta=\pi$.}
\label{fig:invariants_xi_eta}
\end{figure}

Within the FM phase, the metric takes on a simple form:
\be
g_{\xi \xi} = \frac{1}{16} \cot^2 \left( \frac{\eta}{2} \right) ~, ~
g_{\eta \eta} = \frac{1}{16} ~.
\ee
By inspection, the metric and its determinant clearly only diverge at the
anisotropic phase transition, which is at $\eta=0$ in the new parameters.
Within the PM phase, the metric remains quite complicated.
However, one can easily calculate the curvature and determinant
of the metric, which are given by
\begin{widetext}
\beq
\nn
K&=& 8+\frac{8 \csc (\xi)}{\sqrt{-1+\csc^2 (\xi) + \tan^4 \left( \frac{\eta}{2} \right) }} \\
g&=&\csc^2 (\xi) \sec^4 \left( \frac{\eta}{2} \right) \tan^6 \left( \frac{\eta}{2} \right)
\left[ \frac{-1+2\csc^2 (\xi) + \tan^4 \left( \frac{\eta}{2} \right) -2 \csc (\xi)
\sqrt{ -1+\csc^2 (\xi)+\tan^4\left( \frac{\eta}{2}\right)}}
{256 \left( \tan^4 \left( \frac{\eta}{2} \right) - 1\right)^2 
\left( -1 + \csc^2 (\xi) + \tan^4 \left( \frac{\eta}{2} \right) \right)^2} \right]
\eeq
\end{widetext}
These two quantities are plotted for the entire phase diagram in \fref{fig:invariants_xi_eta}.

Clearly we have almost achieved our goal, in that the invariant area component of
the bulk Euler integral, $dS = \sqrt g ~ d\xi d\eta$ is finite except near
a few select points, namely
\begin{itemize}
\item In vicinity of the critical line at $\eta=0$ (i.e. $\gamma=0$), where
the curvature also diverges.
\item Near the points $|h|,|\gamma| \to \infty$, which correspond to 
$|\xi|=|\eta|=\pi$.  Here the curvature does not diverge, so we expect the
divergence of the metric to again be removable.
\end{itemize}

To see how to remove the divergence in $g$ near $\eta=\xi=\pi$,
we need to understand the asymptotics of $g$ near this point.  
We can do leading order asymptotic expansions of the numerator
and denominator about this point. Then if we define 
\be
u\equiv \frac{1}{\csc (\xi)} \approx \pi-\xi ~, ~~~~
v\equiv \frac{1}{\tan^2 (\eta/2)} \approx \left( \frac{\pi - \eta}{2} \right)^2 ~,
\ee
we find that the determinant is asymptotically equivalent to 
\beq
g & \approx & \frac{u^{-2} v^{-5} \big[ 2 u^{-2} + v^{-2} - 2 u^{-1} \sqrt{u^{-2} + v^{-2}}
\big] }{256 v^{-4} \big[ u^{-2} + v^{-2} \big] ^2} \\
& = & \frac{\left( \frac{v}{u} \right) 
\big[ 2 \left(\frac{v}{u}\right)^{2} + 1 - 2 \left(\frac{v}{u}\right) 
\sqrt{1+\left(\frac{v}{u}\right)^2} \big] }{256 u 
\big[ 1 + \left( \frac{v}{u} \right)^2 \big] ^2} ~.
\eeq
Rewriting this in circular coordinates $u=r\cos(\theta)$ and 
$v=r\sin(\theta)$, the expression becomes
$g_{\{\xi , \eta \}}=\frac{1}{256 r} f(\theta)$, where
the notation $g_{\{\xi , \eta \}}$ is meant to reiterate that this
is the determinant of the matrix $\left( \begin{array}{cc} g_{\xi \xi} & g_{\xi \eta} \\
g_{\eta \xi} & g_{\eta \eta} \end{array} \right)$. The function
\be
f(\theta) = \sin \theta \big[ \sin \theta - 1 \big]^2 
\ee
is defined over the interval $\theta \in [0,\pi/2]$.  Then the invariant area is
(using the expansions of $u$ and $v$ from above)
\beq
\nn
dS&=&\sqrt {g_{\{\xi , \eta \}}} d \xi d \eta \\
\nn
&=& \sqrt{\frac{1}{256 r} \sin \theta (1 - \sin \theta)^2}
\Big( -d u \Big) \Big( -\frac{dv}{2 \sqrt v} \Big) \\
\nn
&=& \sqrt{\frac{1}{256 r} \sin \theta (1 - \sin \theta)^2 \left(
\frac{1}{4 r \sin \theta} \right) } du dv \\
\nn
&=& \frac{1 - \sin \theta}{32 r} (r dr d\theta) \\
&=& \frac{1 - \sin \theta}{32}  dr d\theta \equiv \sqrt{g_{\{r,\theta\}}} dr d\theta ~.
\eeq
So we come to our final result that, by choosing a local parameterization
$(r,\theta)$ as described above for the points near $|\xi|=|\eta|=\pi$, 
the metric determinant $g_{\{r,\theta\}}$ is non-divergent.  We
conclude that, after a suitable choice of local reparameterization, the 
invariant area term and its integral can be made finite unless curvature diverges.
Therefore, all divergences in the bulk Euler integral of the $h-\gamma$ plane are due
to the divergent curvature near $\gamma=0$.

\section{Full three-dimensional curvature tensor}
\label{sec:curv_3d}

In this section, we solve for the Riemann curvature tensor, Ricci
tensor, and (Ricci) scalar curvature of the full 3D manifold of the XY
Hamiltonian \eref{eq:H_XY}.  While we remain unable to demonstrate any sharp
physical implications of these tensor components, they do give 
insight geometrically into properties of the 3D Riemann manifold.

\subsection{Ferromagnet} 

For $0<h<1$ and $\gamma>0$ we have for the metric tensor
\beq
\nn g_{\one\one}&=&\frac{1}{16}\frac{1}{\gamma(1+\gamma)^{2}}\\
\nn g_{\two\two}&=&\frac{1}{8}\frac{\gamma}{1+\gamma}\\
g_{\three\three}&=&\frac{1}{16}\frac{1}{\gamma(1-h^{2})}~,
\eeq
where parameters are labeled $h=\lambda_1$, $\gamma=\lambda_2$, and 
$\phi=\lambda_3$.
This is used then to compute the determinant $g=1/[2048(1-h^{2})\gamma(1+\gamma)^{3}]$ and the inverse metric
\beq
\nn g^{\one\one}&=&16\gamma(1+\gamma)^{2}\\
\nn g^{\two\two}&=&8\frac{1+\gamma}{\gamma}\\
g^{\three\three}&=&16\gamma(1-h^{2})
\eeq
The non-zero components of the Christoffel symbols are
\beq
\nn \Gamma^{\two}_{\two\one} &=&\Gamma^{\two}_{\one\two}=\frac{1}{2\gamma(1+\gamma)}, \qquad \Gamma^{\three}_{\three\three}=
\frac{h}{1-h^{2}},\\
\nn \Gamma^{\three}_{\three\one}&=&\Gamma^{\three}_{\one\three}=-\frac{1}{2\gamma},\qquad \Gamma^{\one}_{\two\two}=-\gamma\\
\Gamma^{\one}_{\three\three}&=&\frac{(1+\gamma)^{2}}{2\gamma(1-h^{2})},\qquad \Gamma^{\one}_{\one\one}=
-\frac{1+3\gamma}{2\gamma(1+\gamma)} ~.
\eeq
The nonzero components of the Riemann tensor 
$R^{\rho}_{\sigma\mu\nu}=\partial_{\mu}\Gamma^{\rho}_{\nu\sigma}-
\partial_{\nu}\Gamma^{\rho}_{\mu\sigma}+\Gamma^{\rho}_{\mu\lambda}\Gamma^{\lambda}_{\nu\sigma}-
\Gamma^{\rho}_{\nu\lambda}\Gamma^{\lambda}_{\mu\sigma}$ are
\beq
\nn R^{\two}_{\three\two\three}&=&-R^{\two}_{\three\three\two}=\frac{1+\gamma}{4\gamma^{2}(1-h^{2})},\\
\nn R^{\two}_{\one\two\one}&=&-R^{\two}_{\one\one\two}=\frac{1}{4\gamma(1+\gamma)^{2}}\\
\nn R^{\three}_{\two\three\two}&=&-R^{\three}_{\two\two\three}=\frac{1}{2}\\
\nn R^{\three}_{\one\one\three}&=&-R^{\three}_{\one\three\one}=\frac{1}{2\gamma^{2}(1+\gamma)}\\
\nn R^{\one}_{\two\one\two}&=&-R^{\one}_{\two\two\one}=\frac{\gamma}{2(1+\gamma)}\\
R^{\one}_{\three\three\one}&=&-R^{\one}_{\three\one\three}=\frac{1+\gamma}{2\gamma^{2}(1-h^{2})} ~.
\eeq
The nonzero components of the Ricci tensor $R_{\sigma\nu}=R^{\rho}_{\sigma\rho\nu}$ are
\beq
\nn R_{\one\one}&=&\frac{1}{4\gamma(1+\gamma)^{2}}-\frac{1}{2\gamma^{2}(1+\gamma)} \\
\nn R_{\two\two}&=&\frac{1}{2}+\frac{\gamma}{2(1+\gamma)}\\
R_{\three\three}&=&-\frac{1+\gamma}{4\gamma^{2}(1-h^{2})} ~.
\eeq
The scalar curvature is obtained by contracting $R^{\mu}_{\nu}=g^{\mu\sigma}R_{\sigma\nu}$ 
to get $R=R^{\mu}_{\mu}$. We therefore obtain 
\beq
R=-\frac{8}{\gamma}.
\eeq
This and previous information can be used to compute the Einstein tensor 
$G_{ij}=R_{ij}-\frac{1}{2}g_{ij}R$, which in our case has the following non=zero components
\beq
\nn G_{\one\one}&=&-\frac{1}{4\gamma^{2}(1+\gamma)}~,~G_{\two\two}=1\\
\qquad G_{\three\three}&=&-\frac{1}{4\gamma(1-h^{2})} ~.
\eeq

\subsection{Paramagnet}

For the paramagnet ($\gamma>0$ and $h > 1$), the metric is no longer diagonal, although it
is block diagonal with the form
\be
g = \left( \begin{array}{cc} g_{\{ 1,2 \}} & 0 \\ 0 & g_{33} \end{array} \right) ~.
\ee
The inverse metric tensor and Ricci tensor have this same block diagonal form.
Their expressions are generally quite complicated, so we will not reproduce them here.
However, they can be contracted to give a fairly simple form for the Ricci (1,1) tensor $R^\mu_\nu$,
which has non-zero components
\begin{widetext}
\beq
\nn R^\one_\one &=& \frac{1}{\gamma^2} ~ \frac{4 (2 - 2 h^2 + 3 \gamma^2) 
(h + \sqrt{h^2 + \gamma^2 - 1})}{\sqrt{h^2 + \gamma^2 - 1}}
\\\nn R^\one_\two &=& -\frac{1}{\gamma} ~ \frac{8 (h^2 - 1)}{\sqrt{h^2 + \gamma^2 - 1}}
\\\nn R^\two_\one &=& -\frac{1}{\gamma} ~ \frac{4(\gamma^2 + 2 h^2 - 1 + 2h
\sqrt{h^2 + \gamma^2 - 1})}{\sqrt{h^2 + \gamma^2 - 1}} 
\\\nn R^\two_\two &=& 12 + \frac{8 h}{\sqrt{h^2 + \gamma^2 - 1}}
\\R^\three_\three &=& 8 + \frac{12 h}{\sqrt{h^2 + \gamma^2 - 1}} - 
\frac{8 (h^2 - 1 + h \sqrt{h^2 + \gamma^2 - 1})}{\gamma^2} ~.
\eeq
The trace of $R^\mu_\nu$ gives the scalar curvature:
\be
R = 8\left[ 4 + \frac{5h}{\sqrt{h^2 + \gamma^2 - 1}} - 2 \frac{\left(h^2 + h \sqrt{h^2 + \gamma^2
- 1} - 1 \right)}{\gamma^2} \right] ~.
\ee
\end{widetext}

\begin{figure}
\includegraphics[width=\linewidth]{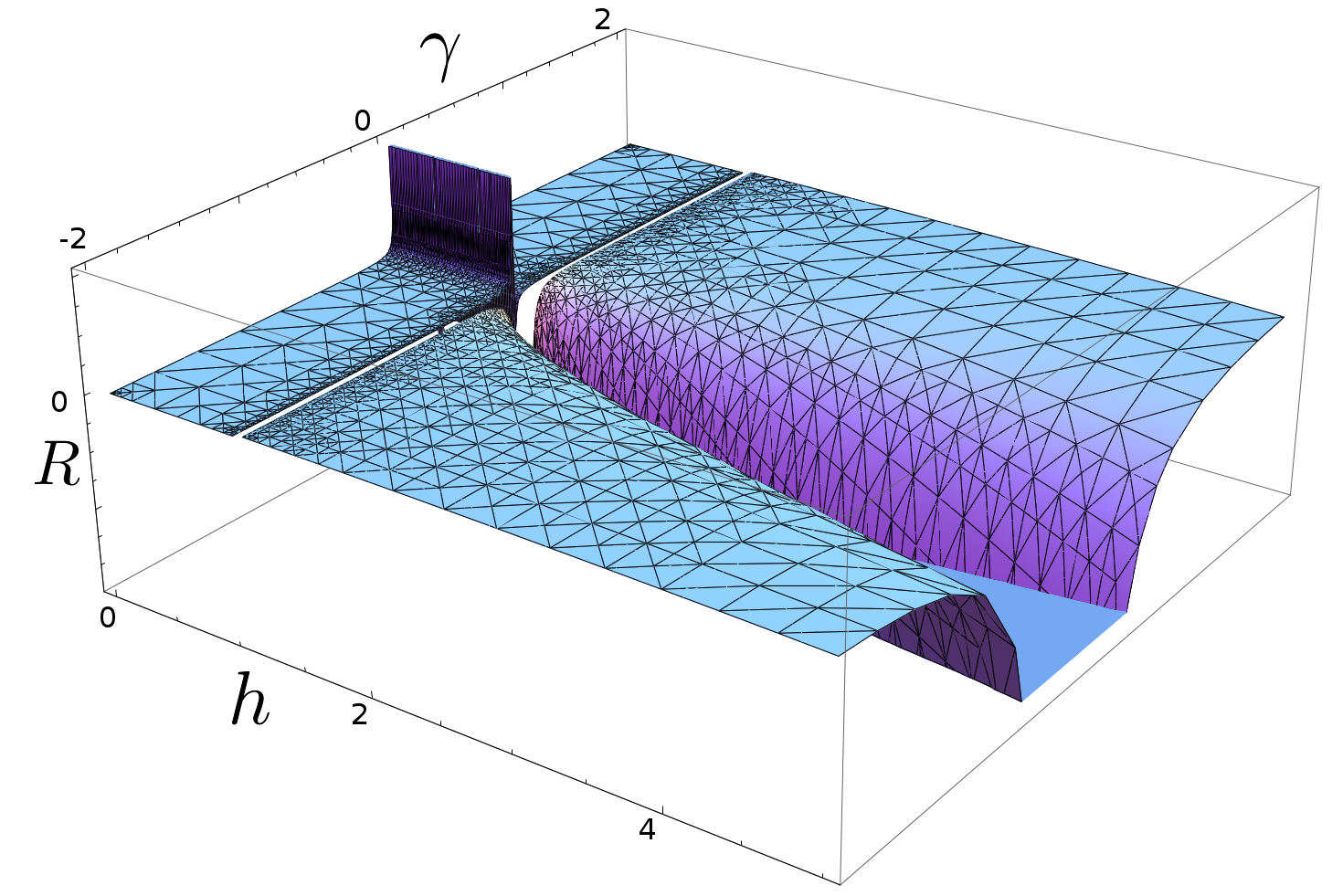}
\caption{Three dimensional scalar curvature $R$ as a function
of $h$ and $\gamma$.}
\label{fig:xy_ricci_curvature}
\end{figure}

Unlike the two-dimensional curvature in the $h-\gamma$ plane, the three dimensional
scalar curvature has divergences far from any critical points.
For instance, at $h \gg 1$, $R \sim h^2$, so it diverges in this limit.
Similarly, the scalar curvature diverges near the line of XY symmetry,
$R \sim 1/\gamma^2$.  We note that, similar to the 3D Ricci scalar $R$,
the 2D Gauss curvature of the $h-\phi$ plane also diverges 
as $K\sim 1/\gamma^2$.  Geometrically, we are not aware of many
results regarding three dimensional manifolds with divergent (negative) 
scalar curvature, but that is indeed what occurs near the line of XY symmetry.
Importantly, this is associated with a singular metric, in the sense that
both $g_{hh}$ and $g_{\phi\phi}$ vanish at $\gamma=0$.  While these divergences
are quite interesting and merit further exploration, we have been unable to draw any
further physical or geometrical conclusions about them at this time.

\bibliography{../ref/References}

\end{document}